\documentclass[final,3p,times]{elsarticle}

\usepackage{geometry}
\usepackage{graphicx}
\usepackage{amssymb}
\usepackage{booktabs}

\usepackage{caption}
\usepackage{subcaption}
\usepackage{lineno}
\usepackage{array}
\usepackage{float}
\usepackage{tcolorbox}
\usepackage{amsmath,amsthm}
\usepackage{mathrsfs}
\usepackage[colorlinks=true, allcolors=blue]{hyperref}
\usepackage[nodots]{numcompress}
\usepackage{arydshln}

\biboptions{sort&compress}
\theoremstyle{remark}

\journal{arXiv}

\begin{document}
\begin{frontmatter}

\title{An Inelastic Homogenization Framework for Layered Materials with Planes of Weakness}

\author[label1]{Shabnam J. Semnani\corref{cor1}}
\ead{ssemnani@ucsd.edu}

\author[label2]{Joshua A. White}
\ead{jawhite@llnl.gov}

\address[label1]{Department of Structural Engineering, University of California, San Diego, United States}

\address[label2]{Atmospheric, Earth and Energy Division,
                 Lawrence Livermore National Laboratory,
                 United States }

\cortext[cor1]{Corresponding author}

\begin{abstract}
Many geologic materials have a composite structure, in which macroscopic mechanical behavior is determined by the properties, shape, and heterogeneous distribution of individual constituents. In particular, sedimentary rocks commonly exhibit a layered microstructure, with distinct bedding planes that can also form planes of weakness. In this work, we present a homogenization framework for modeling inelastic layered media. The proposed constitutive model allows for distinct micro-constitutive laws for each layer, explicit representation of layer distributions, as well as incorporation of imperfect bonding at the interface between adjacent layers. No \emph{a priori} assumptions are needed regarding the specific consitutive models used for the layers and interfaces, providing significant modeling flexibility.  The overall framework provides a simple and physically-motivated way of defining anisotropic material behavior as an emergent property of the layered microstructure.  The model is calibrated using triaxial and true-triaxial experimental data to demonstrate its ability to describe anisotropic deformation and multiple modes of failure.

\end{abstract}


\end{frontmatter}


\section{Introduction}
\label{S:intro}
Many geologic materials are multi-constituent composites. For example, sedimentary rocks commonly consist of distinct material layers that have been deposited over geologic time.  An interesting question is to what degree the composite material behavior can be predicted from knowledge of the individual constituents. For example, elastic properties of layered rocks \cite{Salamon1968ElasticMass} or soils \cite{Sawicki1978OnSoil} can be readily derived via homogenization of the properties within a representative volume element. Layer-to-layer interactions induce a transversely isotropic material response even if the separate layers are  isotropic.  A more challenging question is how to homogenize behavior in the presence of plasticity and  inelastic deformation.  To date, most constitutive models for plastic behavior of transversely isotropic materials take a single-scale viewpoint, with no explicit representation of the material microstructure (e.g. \cite{Semnani2016ThermoplasticityPlasticity,Pariseau1968PlasticitySoil}).  Rather,  the effect of layering is introduced through anisotropy parameters or fabric tensors postulated at the macroscopic level.

Another common limitation is that many constitutive models ignore debonding and sliding that may occur at layer interfaces.
It is clear that these interfaces can significantly affect the mechanical deformation \cite{Semnani2016ThermoplasticityPlasticity,Tien2001ARocks,Drzal1990TheProperties}. In particular, two main modes of failure for sedimentary rocks include sliding along weak discontinuities and shear localization within matrix layers. To capture these multi-mode failure mechanisms, a variety of failure criteria for transversely isotropic rocks have been proposed over the years \cite{Jaeger1960ShearRocks,Hoek1990EstimatingCriterion,Hoek2002Hoek-BrownEdition,Tien2001ARocks}. These models, however, typically focus on describing just the brittle failure envelope, and do not predict the complete stress-strain response.

In this work, we propose a novel strategy for modeling anisotropic, layered media.  We rely on an asymptoptic homogenization framework, in which the governing momentum and traction balance equations for the layered composite are expanded in a multi-scale representation.  This leads to a system of microscale equations that determine the homogenized stress, strain, and tangent stiffness for the composite.  We present an algorithm to solve these microscale equations that has the same structure as an implicit return-mapping algorithm familiar from classical plasticity theory.  As a result, the proposed model may be readily incorporated in standard finite element analysis software as a material point subroutine with no special modification.  The framework is modular is the sense that arbitrary constitutive models may be adopted for the individual layers and interfaces.  This provides significant flexibility to describe a wide array of layered composite material behavior.

The paper is organized as follows. We first provide a brief overview of the homogenization literature in Section \ref{S:background} to put the current work into context. The multi-scale conceptual model is described in Section \ref{S:problem_statement}, followed by a two-scale asymptotic homogenization of the governing equations in Section \ref{S:homog}. The key result is a numerical material point integration algorithm presented in Section \ref{S:implementation}.  The resulting model is calibrated using triaxial and true-triaxial experimental data for several rock types, and numerical results are provided in Section \ref{S:examples} to demonstrate the capability of the present framework. We end the contribution with a few concluding remarks.

\section{Prior work}\label{S:background}

Numerous homogenization techniques have been developed in the past few decades to establish a connection between microscopic and macroscopic behavior, determine effective properties, and predict macroscopic response of heterogeneous media. Homogenization methods can generally be divided into analytical, semi-analytical and computational techniques, with comprehensive reviews of these methods found in \cite{Kanoute2009MultiscaleReview,Geers2017HomogenizationProblems,Pavliotis2008Multi-scaleHomogenization,OrtolanoGonzalez2013AMaterials}.

The literature on computational homogenizations methods is particularly rich---e.g. FFT-based techniques \cite{Vondrejc2013FFT-basedProject,Moulinec1994AComposites,Michel1999EffectiveApproach}, methods based on transform field analysis \cite{Dvorak1992TransformationMaterials,Marfia2016ComputationalPhenomena,Sepe2013AField,Michel2003NonuniformAnalysis}, the Ritz-Galerkin approach \cite{Wulfinghoff2015EfficientApproach}, and the gradient-enhanced scheme \cite{Kouznetsova2002Multi-scaleScheme}. These methods allow for significant modeling flexibility, though with higher implementation complexity and computational cost.  Analytical and semi-analytical homogenization techniques, on the other hand, are appealing in their simplicity, though the underlying conceptual model must be sufficiently simple to lend itself to analytical analysis.  In general, (semi-)analytical methods can be further sub-divided into two categories. In one category, the material is assumed to be periodic or quasi-periodic, introducing a strong regularity assumption. The method proposed here falls in this category.  The second category addresses cases where the microstructure is not explicitly regular. In this case, one can look to at least two options. The first includes approximate methods, e.g. the orientational averaging method \cite{Lagzdins1992OrientationalSolids}, the self-consistent schemes developed by Hill \cite{Hill1965ContinuumPolycrystals,Hill1965AMATERIALS} and Kr\"{o}ner \cite{Kroner1961ZurVielkristalls} for elastic domains, or effective medium approximations established by Eshelby \cite{Eshelby1957TheProblems}.  The second option also includes methods based on variational principles to determine upper and lower bounds for effective properties \cite{PonteCastaneda1991TheComposites,PonteCastaneda2002Second-orderTheory,Suquet1993OverallComposites}.

 For periodic media, asymptotic homogenization methods \cite{Bakhvalov1989Homogenisation:Media,PentaAnHomogenization,Bensoussan2011AsymptoticStructures,Auriault2010HomogenizationMedia} and mean-field methods \cite{Hori1999OnSolids,Doghri2003HomogenizationAlgorithms,Mori1973AverageInclusions,Mercier2012ComparisonMaterials,Nemat-Nasser2013Micromechanics:Materials,Nemat-Nasser1999AveragingPlasticity,Doghri2003HomogenizationAlgorithms,Perdahcoglu2011ConstitutiveHomogenization} (also referred to as average-field methods) are among the most popular techniques. The asymptotic framework is particularly appealing for the layered composite considered here.  It is based on introducing a fast (or local) coordinate and posing a series expansion of the solution in terms of a scale parameter $\epsilon$. The coefficients of the homogenized equations can then be written in terms of the solution on a unit cell \cite{Kalamkarov2009AsymptoticStructures}. We note that high contrast among the constituent properties may result in a nonlocal structure in the homogenized constitutive relations. This particular case has been addressed in the literature by a number of researchers \cite{Allaire1992HomogenizationConvergence,Zhikov2000OnApplications}.

Methods based on asymptotic expansion were initially developed for linear systems with regular or nearly regular structures, and were referred to as mathematical homogenization \cite{Fish1997ComputationalPractice}. Chung et al. \cite{Chung2001AsymptoticApplications} developed a generic recipe for the asymptotic homogenization approach and its implementation within the Finite Element Method (FEM). Pinho-da-Cruz et al. \cite{Pinho-da-Cruz2009AsymptoticModelling} and Oliveria et al. \cite{Oliveira2009AsymptoticApplications} derived the strong form of asymptotic homogenization for linear elasticity, and discussed its numerical implementation using FEM.
The asymptotic homogenization techniques were later extended to nonlinear problems, e.g. composites with nonlinear constituents \cite{Jansson1992HomogenizedStructure}, nonlinear elastic laminates with imperfect interface contact conditions \cite{Lopez-Realpozo2008EffectiveConditions}, interaction of microcracks \cite{Markenscoff2012AsymptoticMicro-cracks}, layered thermoelectric composites \cite{Yang2013NonlinearComposites}, quasilinear transport equations \cite{Telega2000EffectiveApproximants}, damage evolution \cite{Devries1989HomogenizationStructures,Fish1998COMPUTATIONALHOMOGENIZATION}, and large deformations \cite{Fish2008MathematicalLoading}. Several other researchers have developed advanced variations of the asymptotic homogenization technique \cite{Fish1995Multi-gridCase,Ramirez-Torres2018ThreeTissues,Rezakhani2016AsymptoticMaterials,Li2017AShale,Smyshlyaev2000OnMedia}. Among them, Fish and Belsky \cite{Fish1995MultigridCase,Fish1995Multi-gridCase} developed a variant of this technique that removes the assumption of uniform macroscopic fields within the unit cell, which is advantageous for high-gradient regions of the macroscopic field. Ram\'{i}rez-Torres et al. \cite{Ramirez-Torres2018ThreeTissues} extended asymptotic homogenization to three-scale linear elasticity.  Rezakhani and Cusatis \cite{Rezakhani2016AsymptoticMaterials} combined asymptotic homogenization with  a lattice discrete particle model for quasi-brittle materials to include rotational degrees of freedom. Their method was later applied to modeling of the anisotropic behavior of shale by Li et al. \cite{Li2017AShale}. Higher-order or strain gradient terms in asymptotic expansion homogenization method for elastic media were accounted for in \cite{Smyshlyaev2000OnMedia,Triantafyllidis1996TheModels,Andrianov2008HigherMaterials}.

The majority of the literature on homogenization pertains to elastic materials or plasticity of polycrystals, while problems involving plasticity or dissipation in composites and geomaterials have received less attention. Geological materials often deform beyond the elastic regime, thus limiting the application of multi-scale and homogenization methods to these materials.  The work by Suquet \cite{Suquet1987ElementsMechanics} was one of the first to present a homogenization framework for periodic composites with rigid-plastic and elastic-perfectly plastic constituents. In Suquet's approach, micro-stress and strain fields are written as a sum of an average field and a fluctuating field. In addition, macroscopic internal energy and dissipation (or plastic work) are written as an average of their microscopic counterparts. Suquet \cite{Suquet1987ElementsMechanics} presented the qualitative structure of the macroscopic constitutive law, in which macro-strain and the whole field of plastic strains are considered as the state variables. This leads to an infinite number of internal variables. Therefore, simplifying assumptions were made to derive approximate models such as piecewise constant plastic strains \cite{Suquet1987ElementsMechanics,Pruchnicki1998HomogenizedExpansion}.
Subsequently, a number of researchers built on this general framework  \cite{Pruchnicki1998HomogenizedExpansion,Pruchnicki1994AMaterial,Pruchnicki1998HomogenizedMaterial,Ensan2003ASoils,Lourenco1996AMaterials}.
Among them, Pruchnicki \cite{Pruchnicki1998HomogenizedExpansion} discretized the unit cell into subregions with constant plastic microstrain tensors, and used a Fourier series approach to solve the integral equation.

Construction of macroscopic plastic constitutive equations for arbitrary loading typically requires simplifying assumptions regarding stress or strain fields within the individual phases of composites \cite{Aboudi2003Higher-orderPhases}. The Method of Cells \cite{Aboudi1982AComposites} and Generalized Method of Cells \cite{Paley1992MicromechanicalModel} are analytical methods based on a first-order representation of displacement field in each subcell, thus piece-wise uniform stress and strain fields in the periodic unit cell. The transformation field analysis approach is another method used in the literature to reduce the number of macroscopic internal variables by assuming a piecewise uniform \cite{Dvorak1992TransformationMaterials} or a non-uniform \cite{Michel2003NonuniformAnalysis,Covezzi2017HomogenizationTFA,Sepe2013AField} distribution of microscopic fields of internal variables. Fourier series approximation of stress and strain fields within each unit cell is another strategy adopted by a number of researchers \cite{Fotiu1996OverallComposites,Walker1994ThermoviscoplasticSubvolumes}.
Recently, rigorous mathematical homogenization of plasticity equations has also been presented \cite{Schweizer2011TheHomogenization,Heida2016Non-periodicEquations,Allaire1992HomogenizationConvergence,Zhikov2000OnApplications,Visintin2005OnElasto-plasticity,Schweizer2015HomogenizationMethods,Nesenenko2007HomogenizationViscoplasticity,Francfort2014OnElasto-plasticity,Sab1994HomogenizationPlasticity}.

A number of researchers have extended previously developed homogenization methods to elasto-plasticity problems, e.g. asymptotic homogenization \cite{Fish1997ComputationalPractice,Chung2001AsymptoticApplications,Pruchnicki1998HomogenizedMaterial,Ohno2000HomogenizedStructures,Ramirez-Torres2018AnMedia,Chung2004ALoads}, mean-field methods \cite{Doghri2003HomogenizationAlgorithms,Doghri2016FiniteConstituents,Perdahcoglu2011ConstitutiveHomogenization}, as well as methods based on variational principles \cite{PonteCastaneda2002Second-orderTheory,Danas2008AMedia,Suquet1993OverallComposites,Suquet1997EffectiveComposites,PonteCastaneda1991TheComposites,Suquet1993OverallComposites,Talbot1985VariationalMedia,PonteCastaneda1996ExactMaterials,Agoras2011HomogenizationComposites}. Fish et al. \cite{Fish1997ComputationalPractice} generalized the asymptoptic method by regarding all inelastic strains as eigenstrains in an otherwise elastic domain and approximating displacements and eigenstrains (i.e., plastic strains) by an asymptotic power series expansion. Chatzigeorgiou et al. \cite{Chatzigeorgiou2016PeriodicMaterials} applied asymptotic expansion homogenization to thermo-mechanical coupling of multi-layered dissipative standard generalized materials. Variational asymptotic homogenization was developed by Yu and Tang \cite{Yu2007VariationalMaterials} to combine the advantages of variational methods and asymptotic methods by asymptotic expansion of the energy functional for a unit cell, and this approach was later applied to elasto-plastic problems \cite{Zhang2015VariationalComposites}.

A number of researchers have adopted a different view for homogenization involving plasticity, and focused on deriving the homogenized or effective yield limit \cite{BouchitteHOMOGENIZATIONDESIGN,Gluge2016EffectiveMaterials,Gluge2017EffectiveMaterials,Pruchnicki1994AMaterial,Sawicki1981YieldComposites,DeBuhan1991AMaterials,PonteCastaneda1992OnComposites,DeBotton1992OnMaterials,Shen2017ApproximateVoids}. Among them, Sawicki \cite{Sawicki1981YieldComposites} formulated the effective yield surface for two-phase layered materials; however, they did not consider its evolution due to plastic strains. Gl\"{u}ge \cite{Gluge2016EffectiveMaterials} used stress concentration tensors to obtain effective yield surface and effective plastic flow for two-phase laminates, and showed that the effective yield surface evolves with plastic deformations, even if each layer is considered isotropic and elasto-perfectly plastic. Gl\"{u}ge \cite{Gluge2017EffectiveMaterials} examined the application of orientation averaging to derive homogenized plastic properties of laminates. Ponte Casta\~{n}eda and deBotton \cite{PonteCastaneda1992OnComposites} focused on obtaining bounds and estimates for effective yield strength of rigid-perfectly plastic two-phase composites using a variational approach. Shen et al. \cite{Shen2017ApproximateVoids} derived closed forms of the macroscopic yield criterion for porous materials with Drucker-Prager type plastic matrix and spherical voids.

A particular class of periodic materials are layered microstructures, which is the specific focus of the present framework. A number of researchers have made different simplifying assumption to address plasticity in layered materials \cite{ElOmri2000Elastic-plasticComposites,He2012HomogenizationResults,Poulios2018ADeformations}. Among them, El Omri et al. \cite{ElOmri2000Elastic-plasticComposites} derived a homogenization approach for rigid-plastic and elastic perfectly plastic layered composites, assuming plastic strains are unweighted averages of the microscopic ones. This assumption, however, holds if the multi-layered medium has homogeneous elastic properties \cite{Pruchnicki1994AMaterial,He2012HomogenizationResults}. He and Feng \cite{He2012HomogenizationResults} derived exact closed-form solutions for macroscopic behavior of elastic perfectly plastic periodic composites with perfectly bonded layers. They derived the formulation in the general anisotropic case, and defined the macroscopic elastic energy and plastic dissipation power as volume averages of their microscopic counterparts.

Layered geometry allows for simplifying the general homogenization plasticity framework of Suquet \cite{Suquet1987ElementsMechanics}. It has been shown that for multi-layered media with homogeneous layers, the microscopic stresses and strains are constant in each layer, as a direct result of equilibrium equations, compatibility and constitutive relations \cite{Pruchnicki1994AMaterial,Pruchnicki1998HomogenizedExpansion}. The macroscopic quantities can therefore be written as volume averages of the microscopic quantities, in the absence of localization. Following the approach presented by Suquet \cite{Suquet1987ElementsMechanics}, a number of researchers have assumed a decomposition of the strain field into a macroscopic average term and a fluctuating term, and taken advantage of uniformity of stresses and strains in each layer to develop homogenization procedures for elasto-plastic behavior of layered media \cite{Pruchnicki1994AMaterial,Pruchnicki1998HomogenizedMaterial,Ensan2003ASoils,Lourenco1996AMaterials}. Among them, Louren\c{c}o \cite{Lourenco1996AMaterials} derived a matrix formulation for homogenization of elasto-plastic periodic layered media, assuming a homogeneous state of stress and strain in each layer, piecewise constant inelastic strains, and no relative displacement in the interface between layers. They derived an equivalent return mapping algorithm for the average strains and average stresses, in which the Jacobian is calculated numerically. Pruchnicki and Shahrour \cite{Pruchnicki1994AMaterial} adopted a thermodynamical approach to derive a macroscopic constitutive law for multi-layered media with elastic perfectly plastic constituents. They defined the macroscopic free energy and dual potential as volume averages of their microscopic counterparts, and showed that the macroscopic yield surface undergoes kinematic hardening as a result of microscopic plastic deformations.
The work of Pruchnicki and Shahrour \cite{Pruchnicki1994AMaterial} was later extended by Pruchnicki \cite{Pruchnicki1998HomogenizedMaterial} as well as Ensan and Shahrour \cite{Ensan2003ASoils} to introduce imperfect interfaces which allow for slippage at the interface of layers.

In the present work, we describe how the asymptotic homogenization technique can be extended to address inelasticity and imperfect bonding in layered materials, by developing the formulation in rate form and taking advantage of the layered geometry of the unit cell.
We will show that the same form proposed by Suquet \cite{Suquet1987ElementsMechanics} for the stress and strain fields, i.e., sum of a fluctuating term and the average term, emerges from the asymptotic homogenization approach. Based on the layered microstructure of the material, we develop a representation and solution strategy for the microscopic unknowns. Moreover, a numerical procedure is presented that can accommodate general micro-constitutive laws for layers and interfaces.

\section{Conceptual model}\label{S:problem_statement}
\label{S:formulation}

\subsection{Microstructure}\label{S:conceptual}
\begin{figure}[t]
\centering
\begin{subfigure}{0.6\textwidth}
\includegraphics[width=3.2in]{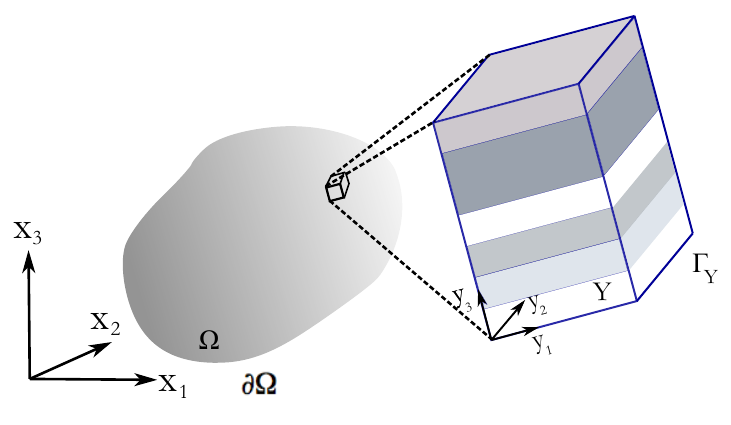}
\caption{}
\label{fig:modela}
\end{subfigure}
\begin{subfigure}{0.3\textwidth}
\centering
\includegraphics[width=1.5in]{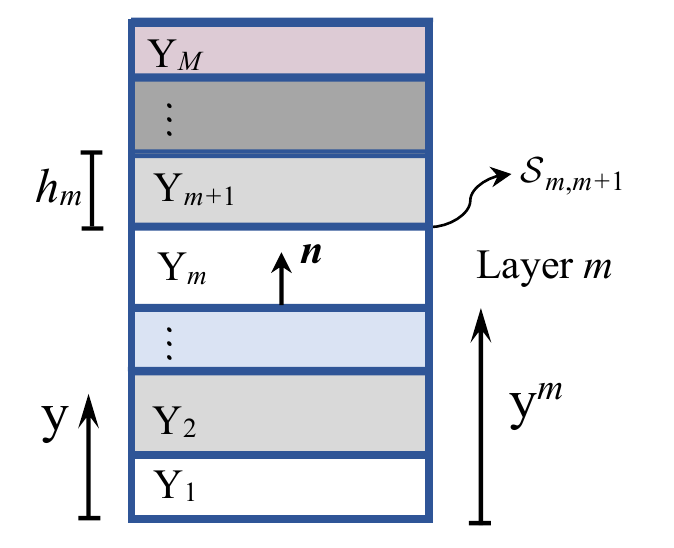}
\caption{}
\label{fig:modelb}
\end{subfigure}
\caption{(a) Schematic representation of unit cell, Y, with multiple parallel homogeneous layers, which compose $\Omega$, the heterogeneous periodic macroscopic medium. $\boldsymbol{n}$ is the unit vector normal to layers. The macroscopic and microscopic coordinate systems are denoted by $\boldsymbol{x}$ and $\boldsymbol{y}$, respectively. (b) Schematic of a unit cell with $M$ layers and local scalar coordinate system $y$. Layer $m$ is shown as $\text{Y}_m$, and $\mathcal{S}_{m,m+1}$ is the interface between layers $m$ and $m+1$.}
\end{figure}

Consider a macroscopic body $\Omega$ composed of a spatially periodic unit cell $Y$ as shown in Figure \ref{fig:modela}. The unit cell consists of parallel layers $Y_m$ with index $m \in L = \{1, 2, ..., M\}$. The material is Y-periodic in the direction normal to the layers, $\boldsymbol{n}$, resulting in a unidirectional microstructure. The surface $\mathcal{S}_{nm}$, identified with double index $nm \in I = \{ 12, 23, ..., M1\}$, separates adjacent layers $n$ and $m$.  Note that from periodicity the final surface $\mathcal{S}_{M1}$ wraps around from the last layer to the first, and the double index proves convenient for tracking such details.  The displacement field is assumed to be continuous within each layer $Y_m$ but may be discontinuous at each interfaces $\mathcal{S}_{nm}$.

Following asymptotic homogenization theory, two separate spatial coordinates are considered: a macroscopic coordinate, $\boldsymbol{x}$, and a microscopic coordinate, $\boldsymbol{y}$. Separation of spatial scales holds, with a non-dimensional period $\epsilon$ defined as
\begin{equation}
\epsilon = \frac{l}{L} \,,\qquad \epsilon \ll 1 \,,
\end{equation}
where $L$ and $l$ denote the macroscopic and microscopic characteristic length scales, respectively. The two coordinate systems are related as
\begin{equation}\label{eq:scale_relation}
\boldsymbol{y}=\boldsymbol{x}/\epsilon \,.
\end{equation}
Due to the uni-directional periodicity of the unit cell, it is convenient to define a scalar micro-coordinate axis, $y$, as (see Figure \ref{fig:modelb})
\begin{equation}\label{eq:micro-coord}
    y = \boldsymbol{y} \cdot \boldsymbol{n}=\frac{1}{\epsilon}\left(\boldsymbol{x} \cdot \boldsymbol{n}\right) \,.
\end{equation}
 The goal of two-scale asymptotic homogenization is to determine the macroscopic behavior of the system when the macroscopic and microscopic scales are sufficiently separate; i.e., for $\epsilon\rightarrow 0$. In this case, it is implicitly assumed in the formulation that the two scales, $\boldsymbol{x}$ and $y$ are independent. As a result, $\boldsymbol{x}$ is treated as a parameter during integration and differentiation with respect to microscopic coordinate $y$.

All material properties are now assumed to be unidirectional, Y-periodic fields that satisfy
\begin{equation}
    f\left( \boldsymbol{x}, y \right)=f\left( \boldsymbol{x}, y + k l \right) \quad \forall k \in \mathbb{Z} \,,
\end{equation}
where $\mathbb{Z}$ is the space of integers. While here we describe a strictly periodic assumption, in practice it is sufficient to have a \emph{quasi-}periodic medium. Material properties may vary with the macroscopic coordinate, provided that the variations with respect to $\boldsymbol{x}$ are much slower than their variations with respect to the microscopic coordinate $y$.  Thus, macroscopic heterogeneities can still be accommodated in the proposed framework. In practice, the material model developed here would be applied at an integration point within a finite element simulation, representing material behavior in a local neighborhood of that point.  Other integration points could be assigned different properties to capture a slow, macroscopic variation of material microstructure and orientation.

\subsection{Governing equations}\label{S:BVP}
Let the domain $\Omega$ be delimited by boundary $\partial \Omega$. The boundary is split as $\partial \Omega = \partial  \Omega_D \cup \partial \Omega_N$, with $\partial \Omega_D \cap \partial \Omega_N = \emptyset$, where subscripts $D$ and $N$ denote Dirichlet and Neumann conditions, respectively.  At the microscale, a periodic set of layers and layer surfaces $\mathcal{S}$ is included as shown in Figure \ref{fig:modelb}.  We introduce a Y-periodic displacement field $\boldsymbol{u}(\boldsymbol{x},y,t)$ with corresponding velocity field $\dot{\boldsymbol{u}}(\boldsymbol{x},y,t)$.   Because we are interested in nonlinear material behavior, it is most convenient to work with the rate form of the governing equations to derive the homogenized model.  Later, we will switch back to a displacement-based formulation for the actual computational implementation.

Within the conceptual geometry, the global initial/boundary value problem is to find $\dot{\boldsymbol{u}}\left(\boldsymbol{x},y,t\right)$ such that
\begin{subequations}
\begin{align}\label{eq:bvpa}
 &\nabla \cdot \dot{\boldsymbol{\sigma}}+\dot{\boldsymbol{f}} =\boldsymbol{0}\qquad &&\text{in } \Omega \setminus {\mathcal{S}} \qquad &&\text{(balance of linear momentum)}\\\label{eq:bvpb}
 & [\![\dot{\boldsymbol{\sigma}} ]\!] \cdot \boldsymbol{n}  = \boldsymbol{0} \qquad &&\text{on } {\mathcal{S}} \qquad &&\text{(traction continuity)} \\\label{eq:bvpe}
& \dot{\boldsymbol{u}} = \dot{{\boldsymbol{u}}}_D  \qquad &&\text{on } \partial  \Omega_D \qquad &&\text{(Dirichlet boundary condition)} \\\label{eq:bvpc}
& \dot{\boldsymbol{\sigma}} \cdot {\boldsymbol{n}_N} = \dot{{\boldsymbol{t}}}_N \qquad &&\text{on } \partial  \Omega_N \qquad &&\text{(Neumann boundary condition)} \\\label{eq:bvpd}
& \boldsymbol{u}\left(t_0\right) = \boldsymbol{u}_0 \qquad &&\text{in} \, \Omega  &&\text{(initial displacement)} \,.
\end{align}
\end{subequations}
The discussion here is limited to quasi-static behavior and infinitesimal deformations.  For brevity, in the remainder of the paper will drop the argument $t$ when specifying the dependent variables of a field, and focus only on the spatial dependencies $\boldsymbol{x}$ and $y$, as they are of primary interest.  It should be clear from the discussion when a field is time-dependent.
%

\subsection{Multiscale kinematics and micro-constitutive models}

\begin{figure}[t]
\centering
\includegraphics[width=1.5in]{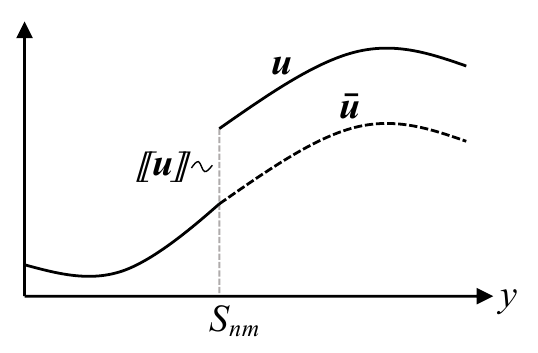}
\caption{Decomposition of the total displacement field $\boldsymbol{u}$  into a continuous part $\bar{\boldsymbol{u}}$ and a jump $[\![ \boldsymbol{u} ]\!]$ in the vicinity of an interface $\mathcal{S}_{nm}$.}
\label{fig:jumpgeo}
\end{figure}

We observe, using Equation (\ref{eq:micro-coord}), that the total derivative of $\boldsymbol{u}(\boldsymbol{x},y)$ with respect to $\boldsymbol{x}$ can be computed as
\begin{equation}\label{eq:derivative}
\frac{\text{d}\boldsymbol{u}}{\text{d}\boldsymbol{x}}=\frac{\partial\boldsymbol{u}}{\partial \boldsymbol{x}}+\frac{1}{\epsilon}\frac{\partial}{\partial y} \left( \boldsymbol{u} \otimes \boldsymbol{n} \right ) \,.
\end{equation}
As a convenient notation, let 
\begin{equation}
\qquad \nabla_x \, {\boldsymbol{u}} = \frac{\partial {\boldsymbol{u}}}{\partial \boldsymbol{x}} \quad \text{and} \quad \nabla_y \,{\boldsymbol{u}} =  \frac{\partial}{\partial y}\left( { \boldsymbol{u}} \otimes \boldsymbol{n}\right)  \,,
\end{equation}
describe macro- and micro-scale gradient operators, respectively.
In our conceptual model, the displacement field within a given layer $Y_m$ is continuous, but it is potentially discontinuous at layer interfaces.  Considering a small neighborhood in the vicinity of a surface $\mathcal{S}_{nm}$ as in Figure~\ref{fig:jumpgeo}, we may express the total displacement field ${ \boldsymbol{u}}$ in the particular form \cite{borja2001, simo1993}
\begin{equation}
{ \boldsymbol{u}}  = \bar{\boldsymbol{u}} + H_{nm}  \, [\![ \boldsymbol{u} ]\!]_{nm}\,.
\end{equation}
Here, $\bar{\boldsymbol{u}} (\boldsymbol{x},y)$ is a continuous field, $H_{nm}(y)$ is a Heaviside function centered at the interface under consideration, and $[\![ \boldsymbol{u} ]\!]_{nm}$ is the displacement jump,
\begin{equation}
[\![ \boldsymbol{u} ]\!]_{nm} = {\boldsymbol{u}}(\boldsymbol{x},y^+)- {\boldsymbol{u}}(\boldsymbol{x},y^-) \,,
\end{equation}
where $y^+$ and $y^-$ denote the micro-coordinates at each side of the discontinuity.  We emphasize that the Heaviside function only depends on the micro-coordinate $y$, as the jump location is determined by the unit cell geometry. 

Noting that the strain field $\hat{\boldsymbol{\varepsilon}}$ is the symmetric gradient of this displacement field, we may write
\begin{equation}\label{eq:strainexp}
\hat{\boldsymbol{\varepsilon}} = \boldsymbol{\varepsilon}_x + \frac{1}{\epsilon} \, \boldsymbol{\varepsilon}_y + \frac{\delta_{nm}}{\epsilon} \,\boldsymbol{\varepsilon}_{nm} \,,
\end{equation}
with $\delta_{nm}$ the Dirac delta function centered at the interface, and strain components
\begin{equation}
\boldsymbol{\varepsilon}_x = \nabla_x^s \,\bar{\boldsymbol{u}} \,, \quad  \boldsymbol{\varepsilon}_y = \frac{\partial}{\partial y} \left(\bar{\boldsymbol{u}} \otimes^s \boldsymbol{n}\right) \,, \quad \text{and} \quad  \boldsymbol{\varepsilon}_{nm} = [\![ \boldsymbol{u} ]\!]_{nm} \otimes^s \boldsymbol{n}  \,.
\end{equation}
Here, $\nabla^s$ and $\otimes^s$ denote symmetric gradient and symmetric dyadic product, respectively.   Due to the appearance of the Dirac delta, the last term in Equation~(\ref{eq:strainexp}) is  singular, but it is only present at the interface itself.  Away from an interface, the strain field is regular and bounded, and can be written as
\begin{equation}\label{eq:regular_eps}
    \boldsymbol{\varepsilon}=\boldsymbol{\varepsilon}_x + \frac{1}{\epsilon} \, \boldsymbol{\varepsilon}_y \,.
\end{equation}

Each layer $Y_m$ in the unit cell is assumed to be homogeneous. The extension of the model to non-homogeneous layers will be discussed in Section \ref{S:multiscale}. We assume the constitutive behavior of individual layers can be expressed in a generic rate form as
\begin{equation}\label{eq:layer_const}
\dot{\boldsymbol{\sigma}}_m =\mathbb{C}_m: \dot{\boldsymbol{\varepsilon}}_m \qquad \forall m \in L \,,
\end{equation}
where $\dot{\boldsymbol{\sigma}}_m$ denotes the stress rate, $\dot{\boldsymbol{\varepsilon}}_m$ the strain rate, and $\mathbb{C}_m$  the
stiffness tangent operator within layer $m$.  This generic form can represent a wide variety of elasto-plastic-damage models for the constituent layers.

The traction rate at an interface is $\dot{\boldsymbol{t}} = \dot{\boldsymbol{\sigma}}\cdot \boldsymbol{n}$.  While the strain rate at an interface may be singular, the traction rate must be bounded in order for the traction equilibrium~(\ref{eq:bvpb}) to be satisfied. To be consistent with Equations~(\ref{eq:strainexp})--(\ref{eq:layer_const}), we assume that the interface constitutive model can be expressed in the  form,
\begin{equation}\label{eq:interface_const}
    \dot{\boldsymbol{t}}_{nm} = \frac{1}{\epsilon} \mathbb{D}_{nm} \cdot  [\![\dot{\boldsymbol{u}} ]\!]_{nm} \,  \qquad \forall nm \in I \,,
\end{equation}
where $\mathbb{D}_{nm}$ is the stiffness tangent operator of the interface.  We will see that the presence of the $1/\epsilon$ weighting here is crucial within the asymptoptic analysis to preserve a proper balancing of terms at various scales in the expansion.  A scale-weighted form is  consistent with traction expressions proposed by other authors in \cite{Pruchnicki1998HomogenizedMaterial,lene1982,shkoller1994,murakami1989}.  This generic form also allows for a variety of specific interface models in order to capture different types of contact behavior. For example, the model in \cite{White2014AnisotropicIntegration} describes a complete framework for describing friction, dilation, and damage on rough interfaces. 

Equations (\ref{eq:layer_const}) and (\ref{eq:interface_const}) let us handle the continuous and discontinuous portions of the displacement field separately, and alleviate the need to directly work with a singular strain field. Therefore, in the remainder of this paper, we will only use the regular part of the strain rate tensor from Equation (\ref{eq:strainexp}). 
It is noteworthy that the homogenized model will now be developed based on the generic micro-constitutive behaviors given in Equations (\ref{eq:layer_const}) and (\ref{eq:interface_const}) without any further specification. We will illustrate the incorporation of specific material models through numerical examples in Section \ref{S:examples}, allowing for a diverse spectrum of macroscopic behavior.

We again remark that in practical applications all fields can be modeled as quasi-periodic; that is, the stiffness tangents $\mathbb{C}$ and $\mathbb{D}$ are functions of both macroscopic and microscopic scales. Their variations at the macroscopic scale are, however, sufficiently slow so that separation is retained. In this case we have
\begin{equation}\label{eq:quasiperiodic}
\frac{\partial\mathbb{C}\left(\boldsymbol{x},y\right)}{\partial \boldsymbol{x}}\ll \frac{\partial\mathbb{C}\left(\boldsymbol{x},y\right)}{\partial y} \quad \text{and} \quad  \frac{\partial\mathbb{D}\left(\boldsymbol{x},y\right)}{\partial \boldsymbol{x}}\ll \frac{\partial\mathbb{D}\left(\boldsymbol{x},y\right)}{\partial y} \,,
\end{equation}
allowing for homogenization on the unit cell. In this case, all response functions, including displacements, stresses, and strains, are also quasi-periodic. 

Before concluding this section, we remark that the divergence of a continuous second order tensor $\boldsymbol{\sigma}$ can be similiarly expanded as
\begin{equation}\label{eq:divergence}
    \nabla  \cdot \boldsymbol{\sigma}  = \nabla_x \cdot \boldsymbol{\sigma} + \frac{1}{\epsilon} \nabla_y \cdot \boldsymbol{\sigma} \,,
\end{equation}
with
\begin{equation}
\label{eq:div_ops}
    \left( \nabla_x  \cdot \boldsymbol{\sigma} \right)_{ij}= \frac{\partial \sigma_{ij}}{\partial x_j} \quad \text{and} \quad \left(\nabla_y  \cdot \boldsymbol{\sigma} \right)_{ij}=  \frac{\partial \sigma_{ij}}{\partial y}  n_j \,.
\end{equation}

\section{Two-scale asymptotic homogenization}\label{S:homog}
Our goal now is to transform the governing formulation into a two-scale system of equilibrium equations, from which the homogenized material behavior may be derived.

\subsection{Series expansions}
Following asymptotic homogenization theory, the unknown velocity field can be represented by an expansion in a power series of $\epsilon$ as
\begin{equation}\label{eq:u_expansion}
 \dot{\boldsymbol{u}}=\sum_{i=0}^\infty \;\epsilon^i \; \dot{\boldsymbol{u}}^{(i)} = \dot{\boldsymbol{u}}^{(0)}+\epsilon \, \dot{\boldsymbol{u}}^{(1)}+\epsilon^2 \, \dot{\boldsymbol{u}}^{(2)} + \mathcal{O}\left(\epsilon^3\right) \,.
\end{equation}
Similarly, the regular portion of the strain rate is expanded as
\begin{align}\label{eq:strain_expansion3}
 \dot{\boldsymbol{\varepsilon}} &= 
\sum_{i=0}^\infty \; \left(\epsilon^i \; \dot{\boldsymbol{\varepsilon}}_x^{(i)} + \epsilon^{i-1} \;\dot{\boldsymbol{\varepsilon}}_y^{(i)} \right) =  \sum_{i=-1}^\infty \;\epsilon^i \; \dot{\boldsymbol{\varepsilon}}^{(i)} \,,
\end{align}
in which we have introduced the notation
\begin{align}
\dot{\boldsymbol{\varepsilon}}^{(-1)}= \dot{\boldsymbol{\varepsilon}}^{(0)}_y \quad \text{and} \quad
 \dot{\boldsymbol{\varepsilon}}^{(i)}=\dot{\boldsymbol{\varepsilon}}_x^{(i)}+\dot{\boldsymbol{\varepsilon}}_y^{(i+1)}  \quad \text{for }i=\{0,1,2,...\} \,.
\end{align}
The stress rate then follows as
\begin{align} \label{eq:stress_expansion}
&\dot{\boldsymbol{\sigma}} = \mathbb{C}:\left( \sum_{i=-1}^\infty \;\epsilon^i \; \dot{\boldsymbol{\varepsilon}}^{(i)}\right) = \sum_{i=-1}^\infty \epsilon^i \; \boldsymbol{\sigma}^{(i)}  \quad \text{with} \quad
\dot{\boldsymbol{\sigma}}^{(i)} =\mathbb{C}:\dot{\boldsymbol{\varepsilon}}^{(i)}  \,.
\end{align}
Similarly, the traction rate is
\begin{equation}\label{eq:traction_rate}
   \dot{\boldsymbol{t}} = \sum_{i=-1}^\infty \;\epsilon^i \;\dot{\boldsymbol{t}}^{(i)}   \quad \text{ with} \quad   \dot{\boldsymbol{t}}^{(i)} = \mathbb{D} \cdot  [\![\dot{\boldsymbol{u}}^{(i+1)}]\!] \,.
\end{equation}
Comparing the stress and traction expansions, we may infer that the relationship $\dot{\boldsymbol{t}}^{(i)} = \dot{\boldsymbol{\sigma}}^{(i)} \cdot \boldsymbol{n}$ also holds.

\subsection{Multiscale equilibrium equations}

Using the macroscopic spatial divergence operator defined in Equation (\ref{eq:divergence}), the balance of linear momentum given in (\ref{eq:bvpa}) can be expanded as
\begin{equation}\label{eq:balance2_momentum}
\nabla_x \cdot \dot{\boldsymbol{\sigma}} + \frac{1}{\epsilon} \nabla_y \cdot \dot{\boldsymbol{\sigma}} + \dot{\boldsymbol{f}} = \boldsymbol{0} \,.
\end{equation}
Replacing the expanded expression for the stress rate tensor given in Equation (\ref{eq:stress_expansion}) into Equation (\ref{eq:balance2_momentum}) leads to
\begin{align}\label{eq:balance}
&\epsilon^{-2} \;\nabla_y \cdot \dot{\boldsymbol{\sigma}}^{(-1)}+\epsilon^{-1}\left(\nabla_x \cdot \dot{\boldsymbol{\sigma}}^{(-1)}+\nabla_y \cdot \dot{\boldsymbol{\sigma}}^{(0)}\right)+\epsilon^0\left(\nabla_x \cdot \dot{\boldsymbol{\sigma}}^{(0)}+\nabla_y \cdot \dot{\boldsymbol{\sigma}}^{(1)} + \dot{\boldsymbol{f}} \right)
+\mathcal{O}\left(\epsilon\right)={\boldsymbol{0}} \,.
\end{align}
Each term multiplied by a power of $\epsilon$ in Equations (\ref{eq:balance})  is now separately set to zero to ensure that the asymptotic series approximation is valid as $\epsilon \to 0$. The results obtained from setting terms of order $\epsilon^{-2}$, $\epsilon^{-1}$, and $\epsilon^0$ to zero imply
\begin{subequations}
\begin{align}\label{eq:equilibriuma}
 & \nabla_y \cdot \dot{\boldsymbol{\sigma}}^{(-1)} = \boldsymbol{0} \,, \\\label{eq:equilibriumb}
& \nabla_x \cdot \dot{\boldsymbol{\sigma}}^{(-1)} + \nabla_y \cdot \dot{\boldsymbol{\sigma}}^{(0)} = \boldsymbol{0} \,, \\\label{eq:equilibriumc}
& \nabla_x \cdot \dot{\boldsymbol{\sigma}}^{(0)}+\nabla_y \cdot \dot{\boldsymbol{\sigma}}^{(1)} + \dot{\boldsymbol{f}}=\boldsymbol{0}  \,.
\end{align}
\end{subequations}
In addition, traction continuity as expressed in Equation (\ref{eq:bvpb}) needs to be satisfied at all scales. Therefore, we have
\begin{equation}\label{eq:traction_expansion}
    [\![\dot{\boldsymbol{\sigma}} ]\!] \cdot \boldsymbol{n}  = \sum_{i=-1}^\infty \; \epsilon^i \, [\![\dot{\boldsymbol{\sigma}}^{(i)} ]\!] \cdot \boldsymbol{n}   = \boldsymbol{0}  \,,
\end{equation}
from which we infer,
\begin{subequations}
\begin{align} \label{eq:BCa}
   & [\![\dot{\boldsymbol{\sigma}}^{(-1)}]\!]  \cdot \boldsymbol{n}  = \boldsymbol{0} \,,\\\label{eq:BCb}
    &[\![\dot{\boldsymbol{\sigma}}^{(0)} ]\!] \cdot \boldsymbol{n}  = \boldsymbol{0} \,.
\end{align}
\end{subequations}
We now examine each of the equilibrium equations in turn.

\subsubsection{Equations (\ref{eq:equilibriuma}) and (\ref{eq:BCa})}
Consider the unit cell $Y$ as shown in Figure \ref{fig:modelb}.
We now multiply both sides of (\ref{eq:equilibriuma}) by $\dot{\boldsymbol{u}}^{(0)}$, integrate over the unit cell, and use the definition of the microscale divergence operator in equation~(\ref{eq:div_ops}), as follows
\begin{align}
0 &= \sum_{m \, \in \, L} \int_{Y_m} \dot{\boldsymbol{u}}^{(0)} \cdot \frac{\partial \dot{\boldsymbol{\sigma}}^{(-1)} }{\partial y} \cdot \boldsymbol{n}  \, \text{d}y \notag \\
&= \sum_{m \, \in \, L} \left[  \dot{\boldsymbol{u}}^{(0)}\cdot \dot{\boldsymbol{\sigma}}^{(-1)} \cdot \boldsymbol{n} \right]^{Y_m^+}_{Y_m^-}
-
\sum_{m \, \in \, L} \int_{Y_m} \frac{\partial  }{\partial y} \left( \dot{\boldsymbol{u}}^{(0)} \otimes \boldsymbol{n} \right) : \dot{\boldsymbol{\sigma}}^{(-1)} \, \text{d}y
\notag \\
&= \sum_{nm \, \in \, I}  [\![ \dot{\boldsymbol{u}}^{(0)} ]\!] \otimes \, \dot{\boldsymbol{t}}^{(-1)}
-
\sum_{m \, \in \, L} \int_{Y_m} \frac{\partial  }{\partial y} \left( \dot{\boldsymbol{u}}^{(0)} \otimes \boldsymbol{n} \right) : \dot{\boldsymbol{\sigma}}^{(-1)} \text{d}y
\notag \\
&= \sum_{nm \, \in \, I}  [\![ \dot{\boldsymbol{u}}^{(0)} ]\!] \cdot \, \mathbb{D} \cdot [\![ \dot{\boldsymbol{u}}^{(0)} ]\!]
-
\sum_{m \, \in \, L} \int_{Y_m} \frac{\partial  }{\partial y} \left( \dot{\boldsymbol{u}}^{(0)} \otimes^s \boldsymbol{n} \right) : \mathbb{C} : \frac{\partial  }{\partial y} \left( \dot{\boldsymbol{u}}^{(0)} \otimes^s \boldsymbol{n} \right) \text{d}y \,.
\end{align}
Both terms must go to zero to satisfy this relationship.  If $\mathbb{C}$ is positive definite, we may infer
\begin{equation}
\frac{\partial^s \dot{\boldsymbol{u}}^{(0)}}{\partial y}=\boldsymbol{0} \qquad \Rightarrow \qquad \dot{\boldsymbol{u}}^{(0)} = \dot{\boldsymbol{u}}^{(0)}\left(\boldsymbol{x}\right) \,.
\end{equation}
That is, $\dot{\boldsymbol{u}}^{(0)}$ is constant in $y$ and \emph{only} depends on $\boldsymbol{x}$. This leads to $\dot{\boldsymbol{\varepsilon}}^{(0)}_y=\boldsymbol{0}$ and consequently, $\dot{\boldsymbol{\sigma}}^{(-1)}=\boldsymbol{0}$. Similarly, if $\mathbb{D}$ is positive definite, we have $[\![ \dot{\boldsymbol{u}}^{(0)} ]\!]=\boldsymbol{0}$ . Therefore, the medium is continuous at the macroscale (as desired) and we only need to consider here velocity jumps on the microscale surfaces in $\mathcal{S}$. 

It should be noted that the assumptions regarding positive definiteness of $\mathbb{C}$ and $\mathbb{D}$ hold in the absence of localization, when separation of scales holds. Slip may still occur at the microscopic interfaces, but in a pervasive and periodic manner.  Of course, a softening mode is likely to be unstable and may eventually localize to one surface in a non-periodic manner.  Such a surface would then need to be explicitly included within the macroscopic  boundary-value-problem ($[\![ \dot{\boldsymbol{u}}^{(0)} ]\!]\neq\boldsymbol{0}$) since a separation-of-scales violation has taken place.

\subsubsection{Equations (\ref{eq:equilibriumb}) and (\ref{eq:BCb})}

Applying the result of equilibrium equation obtained in the previous section to Equation (\ref{eq:equilibriumb}) and using the divergence operator defined for scalar axis $y$ in Equation (\ref{eq:divergence}), we can write
\begin{equation}
\nabla_y \cdot \dot{\boldsymbol{\sigma}}^{(0)} = \frac{\partial}{\partial y} \left(\dot{\boldsymbol{\sigma}}^{(0)} \cdot \boldsymbol{n}\right)= \boldsymbol{0} \,.
\end{equation}
Therefore, $\dot{\boldsymbol{\sigma}}^{(0)} \cdot \boldsymbol{n} = \dot{\boldsymbol{t}}^{(0)}\left(\boldsymbol{x}\right)$ is constant in unit cell $Y$. As a result, the traction continuity condition in Equation (\ref{eq:BCb}) is trivially satisfied.
Consequently, we obtain the following microscale balance equation for each subdomain $Y_m$
\begin{equation}\label{eq:micro_balanceA}
    \left(\mathbb{C}_m:\dot{\boldsymbol{\varepsilon}}^{(0)}\right)\cdot \boldsymbol{n} = \dot{\boldsymbol{t}}^{(0)}\left(\boldsymbol{x}\right) \qquad \forall m \in L \,.
\end{equation}
The equilibrium at interface $\mathcal{S}_{nm}$ is similarly obtained from Equation (\ref{eq:traction_rate}) as
\begin{equation}\label{eq:surface_balance}
    \mathbb{D}_{nm} \cdot [\![\dot{\boldsymbol{u}}^{(1)} ]\!]_{nm} = \dot{\boldsymbol{t}}^{(0)}\left(\boldsymbol{x}\right) \qquad \forall nm \in I \,.
\end{equation}

In Section \ref{S:bi-phasic} we will use the solution of the microscale balance equations (\ref{eq:micro_balanceA}) and (\ref{eq:surface_balance}) to determine $\dot{\boldsymbol{\sigma}}^{(0)}$.

\subsubsection{Equilibrium equation (\ref{eq:equilibriumc})}

Let us define the averaging operator over the unit cell,
\begin{equation}\label{eq:average}
\left<\,\bullet\,\right> = \frac{1}{|Y|}\int_Y\left(\bullet\right) \, \text{d}y \,,
\end{equation}
in which $|Y|$ is volume of the cell.
Applying the averaging operator to both sides of Equation (\ref{eq:equilibriumc}) gives
\begin{equation}\label{eq:ave_eps0_coef}
\frac{1}{|Y|}\left[\int_Y \nabla_x \cdot \dot{\boldsymbol{\sigma}}^{(0)} \, \text{d}y + \int_Y \nabla_y \cdot \dot{\boldsymbol{\sigma}}^{(1)} \, \text{d}y+ \int_Y \dot{\boldsymbol{f}} \, \text{d}y \right] = \boldsymbol{0} \,.
\end{equation}
Here, $\boldsymbol{x}$ is treated as a parameter during integration with respect to microscopic coordinate $\boldsymbol{y}$. Applying the divergence theorem to the second term on the left hand side of Equation ($\ref{eq:ave_eps0_coef}$) leads to:
\begin{equation}\label{eq:eps_0}
\frac{1}{|Y|} \left[ \nabla_x \cdot \int_Y \dot{\boldsymbol{\sigma}}^{(0)}\, \text{d}y + \int_{\Gamma_Y}\dot{\boldsymbol{\sigma}}^{(1)} \cdot \boldsymbol{n} \; \text{d}\Gamma +\int_Y \dot{\boldsymbol{f}} \, \text{d}y \right]=\boldsymbol{0} \,.
\end{equation}
Due to periodicity of the unit cell, the second integral on the left hand side of (\ref{eq:eps_0}) vanishes, and we obtain the macroscopic balance equation
\begin{equation}
\nabla_x \cdot \dot{\boldsymbol{\Sigma}} + \dot{\boldsymbol{F}} = \boldsymbol{0} \qquad \text{in} \qquad \Omega \,.
\end{equation}
Here, the macroscopic stress and body force rates are
\begin{equation}
\dot{\boldsymbol{\Sigma}} \left(\boldsymbol{x}\right)= \left< \dot{\boldsymbol{\sigma}}^{(0)} \right>
\qquad \text{and} \qquad
\dot{\boldsymbol{F}} = \left<\dot{\boldsymbol{f}} \right>  \,.
\end{equation}

\subsection{Two-scale approximation in displacement form}
\label{S:two-scale-approx}
The previous developments have employed the rate form of the governing equations, which simplifies the presentation in the nonlinear material model context.  In practice, however, we often prefer a non-rate, displacement-based form for implementation purposes.  The two are essentially equivalent, so we may readily switch back and forth.

As mentioned earlier, we are interested in  the solution to the problem as $\epsilon \rightarrow 0$. A two-scale expansion is sufficient to provide an approximate solution to the initial/boundary value problem, with an error of $\mathcal{O}(\epsilon)$.  Let us therefore write the displacement field as,
\begin{equation}
{\boldsymbol{u}} (\boldsymbol{x},y) \approx {\boldsymbol{u}}^{(0)} (\boldsymbol{x}) + \epsilon \, {\boldsymbol{u}}^{(1)}  (\boldsymbol{x},y)\,.
\end{equation}
The field ${\boldsymbol{u}}^{(0)}$ provides the macroscopic displacement field in $\Omega$, while ${\boldsymbol{u}}^{(1)}$ provides a microscale ``fluctuation'' that is periodic in $Y$.  The strain  field ${\boldsymbol{\varepsilon}}^{(0)}$ is decomposed as
\begin{align}\label{eq:v_1}
{\boldsymbol{\varepsilon}}^{(0)} = {\boldsymbol{E}}  + {\boldsymbol{e}}  \,,
\end{align}
with macroscale and microscale strain tensors identified, respectively, as
\begin{equation}\label{eq:v_3}
{\boldsymbol{E}}=\nabla_x^s{\boldsymbol{u}}^{(0)} \quad \text{and} \quad {\boldsymbol{e}}=\nabla_y^s {\boldsymbol{u}}^{(1)} \,.
\end{equation}
The two-scale initial/boundary value problem is then:
\begin{itemize}
\item \textit{Macroscale Problem}: Find ${\boldsymbol{u}}^{(0)}\left(\boldsymbol{x}\right)$ such that the following macroscopic conditions are satisfied:
\begin{subequations}
\begin{align}\label{eq:bvpa_macro}
 &\nabla_x \cdot {\boldsymbol{\Sigma}}+{\boldsymbol{F}} =\boldsymbol{0} &&\text{in } \Omega \qquad &&\text{(balance of linear momentum)}\\\label{eq:bvpb_macro}
& {\boldsymbol{u}}^{(0)}= {{\boldsymbol{u}}}_D \qquad &&\text{on } \partial \, \Omega_D  &&\text{(Dirichlet B.C.)} \\\label{eq:bvpc_macro}
& {\boldsymbol{\Sigma}} \cdot {\boldsymbol{n}}_N = {{\boldsymbol{t}}}_N  &&\text{on } \partial \, \Omega_N &&\text{(Neumann B.C.)} \\ \label{eq:bvpe_macro}
& \boldsymbol{u}^{(0)}\left( t_0 \right) = \boldsymbol{u}_0  \qquad &&\text{in} \, \Omega &&\text{(initial displacement)}
\end{align}
\end{subequations}
where the macroscopic stress ${\boldsymbol{\Sigma}} = \left< {\boldsymbol{\sigma}}^{(0)} \right> $ at point $\boldsymbol{x}$ satisfies the microscale problem:

\item \textit{Microscale Problem}: Given ${\boldsymbol{u}}^{(0)}\left(\boldsymbol{x}\right)$, find ${\boldsymbol{u}}^{(1)}\left(\boldsymbol{x},y\right)$ such that
\begin{subequations}
\begin{align}
&{\boldsymbol{\sigma}}^{(0)}   \cdot \boldsymbol{n} = {\boldsymbol{t}}^{(0)}\left(\boldsymbol{x} \right) &&\text{in  } Y_m \; &&\text{for each layer }m \in L \label{eq:micro-balance1}\\
&\boldsymbol{t}^{(0)}_{nm} = {\boldsymbol{t}}^{(0)}\left(\boldsymbol{x}\right) &&\text{on } \mathcal{S}_{nm} \; &&\text{for each interface }nm \in I 
\end{align}
Periodicity conditions requires that:
\begin{align}\label{eq:periodic_BC}
    &{\boldsymbol{u}}^{(1)} \; \text{is periodic in }Y   \\
    &{\boldsymbol{\sigma}}^{(0)} \cdot \boldsymbol{n} \; \text{is anti-periodic on } \Gamma_Y
\end{align}
\end{subequations}
\end{itemize}
The macroscale problem is identical to a typical solid mechanics formulation solved via, e.g., a finite element discretization.  The constitutive relationship between the macroscopic stress  ${\boldsymbol{\Sigma}}$ and macroscopic strain  ${\boldsymbol{E}}$ at an integration point, however, is computed by solving the microscale problem.  Given a macroscopic field ${\boldsymbol{u}}^{(0)}$ with associated strain  ${\boldsymbol{E}}$, we solve the micro-equilibrium equations to determine ${\boldsymbol{\sigma}}^{(0)}$, from which the macroscopic stress  ${\boldsymbol{\Sigma}}$ may be computed by averaging over $Y$.  A procedure for doing so is detailed in the next section.

We remark that for a unit cell located at the macroscopic point $\boldsymbol{x}$, Equation (\ref{eq:v_1}) can be used to obtain the approximate displacement vector ${\boldsymbol{u}}$ at the microscale. For this purpose, Equation (\ref{eq:v_1}) is integrated in terms of $\boldsymbol{y}$, which leads to:
%
%
\begin{equation}\label{eq:v_2}
     {\boldsymbol{u}} \approx {\boldsymbol{E}}\left(\boldsymbol{x}\right) \cdot \boldsymbol{y} + {\boldsymbol{u}}^{(1)} \, .
\end{equation}
Equations (\ref{eq:v_1}-\ref{eq:v_2}) link the asymptotic homogenization approach to a group of other methods used in the literature, for example \cite{Suquet1987ElementsMechanics,He2012HomogenizationResults}, in which the microstrain field is written as sum of an average value and a periodic fluctuation.

\subsection{Solution to the microscale problem}\label{S:bi-phasic}

\begin{figure}[t]
\centering
\includegraphics[width=3.5in]{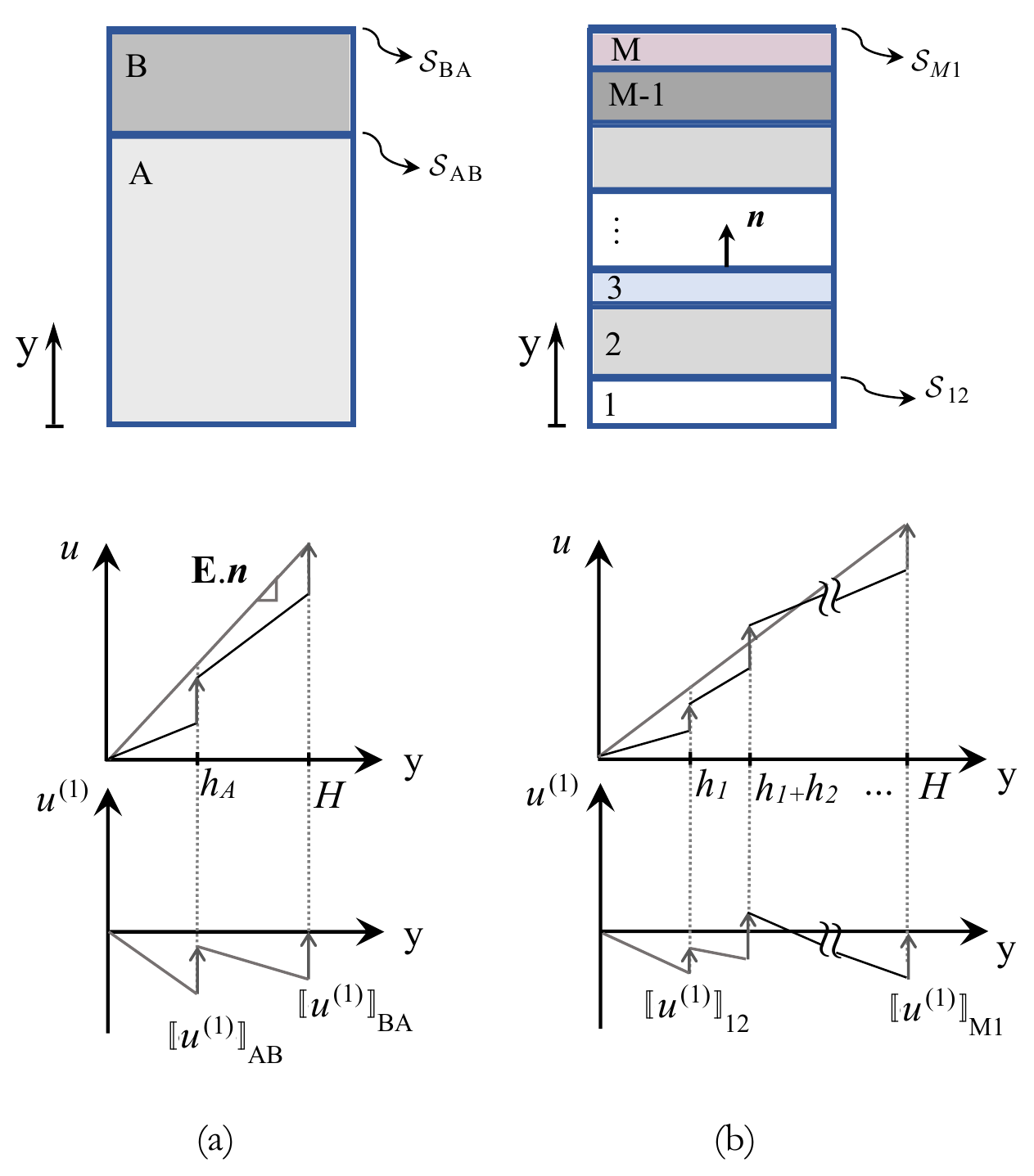}
\caption{Schematic of the total displacement field $\boldsymbol{u}$ and the periodic fluctuation $\boldsymbol{u}^{(1)}$ in (a) bi-layer and (b) multi-layer unit cell.}
\label{fig:unitcell}
\end{figure}

Within the material point update, the macroscale strain $\boldsymbol{E}$ is viewed as a known input, from which we seek to derive the macroscale stress $\boldsymbol{\Sigma}$ and the consistent tangent operator $\overline{\mathbb{C}}$. The general formulation is derived for a multi-layer unit cell with $M$ layers, as shown in Figure \ref{fig:unitcell}. To clarify the formulation further, specific forms are also presented for the simple subcase of a bi-layer unit cell consisting of two parallel layers, $A$ and $B$, and two interfaces, $\mathcal{S}_{AB}$ and $\mathcal{S}_{BA}$.

The volumetric ratio for each phase, $m$, is defined as
\begin{equation}
    \phi_m = \frac{h_m}{l} \quad \text{with} \quad \sum_{m \,\in \,L}\phi_m=1 \,,
\end{equation}
where $l$ is the total thickness of the unit cell.
If layers are individually homogeneous, microscopic strains and stresses are uniform in each layer, as a direct result of equilibrium equations, compatibility, and constitutive relations  \cite{Pruchnicki1994AMaterial,Pruchnicki1998HomogenizedExpansion}.  We will indicate by $\boldsymbol{\varepsilon}_m$ and $\boldsymbol{\sigma}_m$ the restriction of the strain field $\boldsymbol{\varepsilon}^{(0)}$ and stress field $\boldsymbol{\sigma}^{(0)}$ to each layer $m$.

Uniformity of the strain tensor ${\boldsymbol{e}}$ in each layer directly indicates that the microscopic periodic displacement  ${\boldsymbol{u}}^{(1)}$ is a piecewise linear function of $y$. To satisfy the periodic conditions given in (\ref{eq:periodic_BC}), micro-displacements must satisfy the following compatibility condition:
\begin{equation}\label{eq:compatibility}
    \sum_{m \,\in \,L}\int_{Y_m} \frac{\partial {\boldsymbol{u}}^{(1)}}{\partial y} \, \text{d}y + \sum_{nm \, \in \, I} \, [\![{\boldsymbol{u}}^{(1)}]\!]_{nm} = \boldsymbol{0} \,.
\end{equation}
See Figure~\ref{fig:unitcell} for a schematic visualization.  It is logical then to define the following primary unknowns:
\begin{align}\label{eq:unknowns}
\notag    & {\boldsymbol{v}}_m = \frac{\partial {\boldsymbol{u}}^{(1)}_m}{\partial y} &&(\text{constant for each layer } m \in L) \,,\\
    &{\boldsymbol{w}}_{nm} =  [\![{\boldsymbol{u}}^{(1)}]\!]_{nm} &&(\text{constant for each interface } nm \in I ) \,.
\end{align}
Within the solution algorithm, we will solve for one displacement gradient vector ${\boldsymbol{v}}_m$ for each layer, and one displacement jump vector ${\boldsymbol{w}}_{nm}$ for each interface. Note, however, that symmetry requires ${\boldsymbol{w}}_{nm} = {\boldsymbol{w}}_{mn}$ if two materials come into contact multiple times due to periodicity, reducing the number of independent unknowns. This is easiest to see in the bi-layer case in Figure~\ref{fig:unitcell}a.  Surfaces $\mathcal{S}_{AB}$ and $\mathcal{S}_{BA}$ both experience the same loading conditions, and therefore $\boldsymbol{w}_{AB} = \boldsymbol{w}_{BA}$.

Replacing (\ref{eq:unknowns}) into (\ref{eq:compatibility}) gives the simplified compatibility condition
\begin{equation}\label{eq:compatibility2}
    \sum_{m\, \in \, L} \phi_m {\boldsymbol{v}}_m + \sum_{nm \,\in \, I}{\boldsymbol{w}}_{nm} = \boldsymbol{0} \,.
\end{equation}
Using these unknowns, the total strain in layer $m$ is
\begin{equation} \label{eq:eA}
 {\boldsymbol{\varepsilon}}_m = {\boldsymbol{E}} + {\boldsymbol{v}}_m \otimes^s \boldsymbol{n}  \,,
\end{equation}
from which the layer stress $\boldsymbol{\sigma}_m (\boldsymbol{\varepsilon}_m)$ can be computed by calling a material model update routine specific to that layer.  Similarly, the traction $\boldsymbol{t}_{nm}(\boldsymbol{w}_{nm})$ can be computed by calling the desired traction model. The micro-scale balance equations to be satisfied are
\begin{align}
   & {\boldsymbol{\sigma}}_m \cdot \boldsymbol{n} -{\boldsymbol{t}^{(0)}}(\boldsymbol{x}) = \boldsymbol{0} && \text{for all layers } m \in L \label{eq:micro-balance2} \,,\\
   & {\boldsymbol{t}}_{nm} -{\boldsymbol{t}^{(0)}}(\boldsymbol{x}) = \boldsymbol{0} && \text{for all interfaces } nm \in \mathcal{I} \label{eq:micro-balance3} \,.
\end{align}
The normal traction $\boldsymbol{t}^{(0)}(\boldsymbol{x})$, which is constant in $Y$, is also unknown and must be determined by the material point algorithm. For simplicity, in this section we set $\boldsymbol{t}=\boldsymbol{t}^{(0)}(\boldsymbol{x})$.  In total, we have $M$ layer balance equations (\ref{eq:micro-balance2}), $N$ surface balance equations (\ref{eq:micro-balance3}), and one compatibility condition (\ref{eq:compatibility2}).  This provides a fully determined system for the $M$ displacement gradients $\boldsymbol{v}_m$, the $N$ displacement jumps $\boldsymbol{w}_{nm} $, and the overall traction $\boldsymbol{t}$.

  In the following, we now switch to Voigt representation of the tensor quantities to clarify how the algorithm is  implemented. We use brackets $\{ \bullet \}$ to indicate a symmetric tensor that has been unrolled into $6\times1$ algebraic vector. From (\ref{eq:eA}), the layer strain can be computed as
\begin{equation}\label{eq:eA_voigt}
   \{{\boldsymbol{\varepsilon}}_m \} = \{ \boldsymbol{E} \} + \boldsymbol{B} {\boldsymbol{v}}_m \,,
\end{equation}
\begin{equation}
        \{{\boldsymbol{\varepsilon}}_m \}= \begin{bmatrix}
      \varepsilon_{11} & \varepsilon_{22} & \varepsilon_{33} & 2\varepsilon_{23} & 2\varepsilon_{13} & 2\varepsilon_{12}  \end{bmatrix}^{\mathsf{T}} \,,
      \end{equation}
      \begin{equation}
 \boldsymbol{v}_m =   \begin{bmatrix}
  v_1 & v_2 & v_3
\end{bmatrix}^{\mathsf{T}} \,.
\end{equation}
Here,  $\boldsymbol{B}$ is a $6\times 3$ operator that depends on the layer normal $\boldsymbol{n}$,
\begin{equation}
\renewcommand\arraystretch{1.1}
\boldsymbol{B}=
   \begin{bmatrix}
n_1 & 0 & 0    \\
0 & n_2 & 0  \\
0 & 0 & n_3    \\
0 & n_3 &n_2    \\
n_3 & 0 & n_1  \\
n_2 & n_1 & 0    \\
\end{bmatrix} \,.
\end{equation}
Its transpose can be used to express balance equation  (\ref{eq:micro-balance2}) as
\begin{align}
   \boldsymbol{B}^{\mathsf{T}} \{{\boldsymbol{\sigma}}_m\}-{\boldsymbol{t}}&=\boldsymbol{0} \label{eq:balance1} \,,
\end{align}
\begin{equation}
        \{{\boldsymbol{\sigma}}_m \}= \begin{bmatrix}
      {\sigma}_{11} & {\sigma}_{22} & {\sigma}_{33} & {\sigma}_{23} & {\sigma}_{13} & {\sigma}_{12}  \end{bmatrix}^{\mathsf{T}} \,.
      \end{equation}
For the general multilayer system, we can assemble the complete set of nonlinear residual equations as
\begin{equation}
\renewcommand\arraystretch{1.1}
\boldsymbol{\boldsymbol{R}}\left( \boldsymbol{X} \right) =
\begin{bmatrix}
\boldsymbol{B}^\mathsf{T} \{ \boldsymbol{\sigma}_1 \} - \boldsymbol{t} \\
\boldsymbol{B}^\mathsf{T} \{ \boldsymbol{\sigma}_2 \} - \boldsymbol{t} \\
\vdots \\
\boldsymbol{B}^\mathsf{T} \{ \boldsymbol{\sigma}_M \} - \boldsymbol{t} \\
\hdashline
\boldsymbol{t}_{12} - \boldsymbol{t} \\
\boldsymbol{t}_{23} - \boldsymbol{t} \\
\vdots \\
\boldsymbol{t}_{M1} - \boldsymbol{t} \\
\hdashline
\sum_{m} \phi_m \boldsymbol{v_m} + \sum_{nm} \boldsymbol{w}_{nm}
\end{bmatrix}
= \boldsymbol{0}
\quad \text{with} \quad
\boldsymbol{X} =
\begin{bmatrix}
\boldsymbol{v}_1 \\
\boldsymbol{v}_2 \\
\vdots \\
\boldsymbol{v}_M \\
\hdashline
\boldsymbol{w}_{12} \\
\boldsymbol{w}_{23} \\
\vdots \\
\boldsymbol{w}_{M1} \\
\hdashline
\boldsymbol{t}
\end{bmatrix} \,.
\end{equation}
Note that any redundant slip unknowns should be condensed out to avoid an ill-posed system.  The solution to these equations can then be found using, e.g., Newton's method.  Let $\boldsymbol{\mathcal{J}} = \partial \boldsymbol{\boldsymbol{R}} / \partial \boldsymbol{X}$ denote the Jacobian matrix that arises from linearization of the nonlinear residual.   Given a current estimate $\boldsymbol{X}^k$ for the solution, an improved estimate can be found by:
\begin{align}
\text{solving} &\qquad \boldsymbol{\mathcal{J}}^k \Delta \boldsymbol{X} = - \boldsymbol{\boldsymbol{R}}^k \,,\\
\text{updating} &\qquad \boldsymbol{X}^{k+1} = \boldsymbol{X}^{k} + \alpha \Delta \boldsymbol{X} \,.
\end{align}
Here, $\alpha \in (0,1]$ is a suitably chosen line-search or backtracking parameter than can be used to increase the robustness of the search when far away from the neighborhood of convergence.  The sequence is repeated until the residual norm drops below a desired tolerance. At convergence, we then obtain the macroscopic stress tensor as the volumetric average:
\begin{equation}\label{eq:Sigma}
\boldsymbol{\Sigma}=\sum_{m \,\in\, L}  \phi_m \, \boldsymbol{\sigma}_m
\end{equation}
It is perhaps easiest to understand the algorithm structure by examining the special case of a bi-layer medium---i.e. with two layers, $A$ and $B$, and two interfaces, $AB$ and $BA$.  We first identify the symmetry $ {\boldsymbol{w}}_{AB} = {\boldsymbol{w}}_{BA}$, so that one slip unknown is made redundant.  The unknowns and residual equations are then
\begin{equation}
\boldsymbol{\boldsymbol{R}}\left(\boldsymbol{X}\right) = \begin{bmatrix} \boldsymbol{B}^{\mathsf{T}}\{\boldsymbol{\sigma}_A\} -\boldsymbol{t}\\
\boldsymbol{B}^{\mathsf{T}}\{\boldsymbol{\sigma}_B\} -\boldsymbol{t}\\
\boldsymbol{t}_{AB} - \boldsymbol{t}\\
\phi_A \boldsymbol{v}_A + \phi_B \boldsymbol{v}_A + 2 \boldsymbol{w}_{AB}
\end{bmatrix} = \boldsymbol{0}
\quad \text{with} \quad
\boldsymbol{X} = \begin{bmatrix}
 \boldsymbol{v}_A \\
 \boldsymbol{v}_B \\
 \boldsymbol{w}_{AB} \\
 \boldsymbol{t}
\end{bmatrix} \,.
\end{equation}
Note the appearance of the $2$ is the compatibility condition, as we have two surfaces with identical slip ${\boldsymbol{w}}_{AB}$.
The Jacobian for this system is
\begin{equation}\label{eq:J}
   \boldsymbol{\mathcal{J}} =\frac{\partial \boldsymbol{\boldsymbol{R}}}{\partial \boldsymbol{X}} = \begin{bmatrix} \boldsymbol{B}^{\mathsf{T}} \boldsymbol{C}_A \boldsymbol{B} &  & & -\boldsymbol{1} \\[0.5em]
        & \boldsymbol{B}^{\mathsf{T}} \boldsymbol{C}_B \boldsymbol{B} & & -\boldsymbol{1}\\[0.5em]
         &  &  \boldsymbol{D}_{AB} & -\boldsymbol{1} \\[0.5em]
         \phi_A \boldsymbol{1} & \phi_B \boldsymbol{1} &\boldsymbol{2} &
         \end{bmatrix} \,,
\end{equation}
in which $\boldsymbol{1}$ denotes a diagonal identity matrix, and $\boldsymbol{2}$ is two times the identity matrix. Here, $\boldsymbol{C}_A$ and $\boldsymbol{C}_B$ denote the $6\times6$ matrix form of \emph{consistent} (or algorithmic) stiffness tangents of layers $A$ and $B$, which should be available from the layer material model subroutines.  Similarly, $\boldsymbol{D}_{AB}$ is the tangent from the traction model. Throughout this section, all tangent operators are consistent operators, rather than the continuum versions, in order to maintain optimal convergence rates.

At convergence, we may derive the homogenized consistent tangent operator,
\begin{equation}
\overline{\boldsymbol{C}} = \frac{\partial \{ \boldsymbol{\Sigma} \} }{\partial \{ \boldsymbol{E} \} } = \sum_{m \,\in\,L}\phi_m\frac{\partial \{ \boldsymbol{\sigma}_m \} }{\partial \{\boldsymbol{E} \} } \,,
\end{equation}
in which
\begin{equation}
\dfrac{\partial \{ \boldsymbol{\sigma}_m \} }{\partial \{\boldsymbol{E} \}} = \dfrac{\partial \{\boldsymbol{\sigma}_m\}}{\partial \{\boldsymbol{E}\}} \Bigg|_{\boldsymbol{X}}+\frac{\partial \{\boldsymbol{\sigma}_m\}}{\partial \boldsymbol{X}} \Bigg|_{\boldsymbol{E}}\dfrac{\partial \boldsymbol{X}}{\partial \{\boldsymbol{E}\}} = \boldsymbol{C}_m+\boldsymbol{C}_m \boldsymbol{B}\frac{\partial \boldsymbol{v}_m}{\partial \{\boldsymbol{E} \}}
\,.
\end{equation}
To compute the final derivative, we examine the converged residual $\boldsymbol{R}\left( \boldsymbol{X}^*\right)=\boldsymbol{0}$, in which $\boldsymbol{X}^*$ denotes the solution to the local iteration. At this configuration,
\begin{equation}\label{eq:dR_deps_k}
\frac{\partial \boldsymbol{\boldsymbol{R}}}{\partial \{\boldsymbol{\boldsymbol{E}}\}}=\frac{\partial \boldsymbol{\boldsymbol{R}}}{\partial \{\boldsymbol{E}\}} \Bigg|_{\boldsymbol{X}}+\frac{\partial \boldsymbol{\boldsymbol{R}}}{\partial \boldsymbol{X}} \Bigg|_{\boldsymbol{E}}\frac{\partial \boldsymbol{X}}{\partial \{\boldsymbol{E}\}} =\boldsymbol{0} \,.
\end{equation}
Let $\boldsymbol{\mathcal{G}}=\boldsymbol{\mathcal{J}}^{-1}$ be the inverse of the local tangent evaluated at $\boldsymbol{X}^*$. Solving (\ref{eq:dR_deps_k}) gives
\begin{equation}
\frac{\partial \boldsymbol{X}}{\partial \{\boldsymbol{E}\}}= -\boldsymbol{\mathcal{G}} \frac{\partial \boldsymbol{\boldsymbol{R}}}{\partial \{\boldsymbol{E}\}}\Bigg|_{\boldsymbol{X}} \,,
\end{equation}
from which we infer
\begin{equation}
\frac{\partial \boldsymbol{v}_m}{\partial \{\boldsymbol{E} \}} = - \sum_{n \, \in \, L} \boldsymbol{\mathcal{G}}_{mn}\boldsymbol{B}^\mathsf{T} \boldsymbol{C}_n \,.
\end{equation}
Here, $\boldsymbol{\mathcal{G}}_{mn}$ denotes the $3\times3$ sub-block of matrix $\boldsymbol{\mathcal{G}}$ whose row and columns indices correspond to the residual equations for layer $m$ and $n$, respectively. Combining the previous equations, the homogenized tangent operator in matrix form is finally obtained as,
\begin{equation}\label{eq:macro_C}
\overline{\boldsymbol{C}} = \sum_{m \, \in \, L} \phi_m\boldsymbol{C}_m - \sum_{m \, \in \, L} \sum_{n \, \in \, L} \phi_m \boldsymbol{C}_m \boldsymbol{B} \boldsymbol{\mathcal{G}}_{mn} \boldsymbol{B}^{\mathsf{T}}\boldsymbol{C}_n
\end{equation}
It is interesting to observe that this operator is composed of a volume weighted average term plus an inter-layer interaction term.  This latter effect induces anisotropic behavior, even if the layers are isotropic materials.

\subsection{Material point algorithm}\label{S:implementation}
The final material point update is summarized in Algorithm 1.  The state is fully known at time $t^{n-1}$, and the goal is to compute an updated state at time $t^{n}$ given a new macroscopic strain $\boldsymbol{E}^n$ as input. We use $\boldsymbol{\kappa}_m^{n-1}$ and $\boldsymbol{\kappa}_{nm}^{n-1}$ to denote, respectively, internal state parameters for layer $m$ and interface $nm$ from the previous timestep.  Note that this routine can be coded in such a way that the micro-constitutive models are called in a polymorphic fashion, so that one homogenization routine can employ a variety of micro-constitutive models on demand. \\

\begin{tcolorbox}
{\bf Algorithm 1.} Material point update. \\[-0.75em]
\footnotesize

Input: $\boldsymbol{E}^n$

Output: $\boldsymbol{\Sigma}^n$ and $\overline{\boldsymbol{C}}^{\,n}$
\begin{enumerate}
    \item Set local iteration counter $k=0$ and initial guess $\boldsymbol{X}^k=\boldsymbol{0}$ .

         \item For each layer $m \in L$:

    \begin{itemize}
    \item  From $\boldsymbol{X}^k$, compute micro-strain $ \boldsymbol{e}_m$.

    \item Call material subroutine for layer $m$:

    Inputs: $\boldsymbol{\varepsilon}_m = \left(\boldsymbol{E}^n+\boldsymbol{e}_m \right)$, $\boldsymbol{\varepsilon}_m^{n-1}$, $\boldsymbol{\kappa}_m^{n-1}$

    Outputs: $\boldsymbol{\sigma}_{m}$, $\boldsymbol{C}_{m}$, $\boldsymbol{\kappa}_m$

    \end{itemize}

    \item For each interface $nm \in I$:

    \begin{itemize}
    \item  From $\boldsymbol{X}^k$, compute jump $\boldsymbol{w}_{nm}$.

    \item Call material subroutine for interface $nm$:

    Inputs: $ \boldsymbol{w}_{nm}$, $\boldsymbol{w}^{n-1}_{nm}$, $\boldsymbol{\kappa}_{nm}^{n-1}$

    Outputs: $\boldsymbol{t}_{nm}$, $\boldsymbol{D}_{nm}$, $\boldsymbol{\kappa}_{nm}$

    \end{itemize}

    \item Assemble residual $\boldsymbol{\boldsymbol{R}}^k \left(\boldsymbol{X}^k\right)$

     \item If $\|\boldsymbol{\boldsymbol{R}}^k\| < \epsilon_\text{ tol}$ go to Step 8.

    \item Else, perform Newton update $\boldsymbol{X}^{k+1} = \boldsymbol{X}^k - \alpha \left(\boldsymbol{\mathcal{J}}^{-1}\right)^k \, \boldsymbol{\boldsymbol{R}}^k$.

    \item Set $k \leftarrow k+1$ and return to Step 2.
      \item Compute the macro-stress $\boldsymbol{\Sigma}^{n} = \sum_{m} \phi_m\boldsymbol{\sigma}_m$
    \item Compute the consistent tangent operator $\overline{\boldsymbol{C}}^{\,n}$ via equation~(\ref{eq:macro_C}).

      \item Save the sub-model states in preparation for the next timestep:
    \begin{itemize}
        \item For each layer: $\boldsymbol{\varepsilon}_m^{n} \leftarrow \boldsymbol{\varepsilon}_m$,
        $\boldsymbol{\kappa}_m^{n} \leftarrow \boldsymbol{\kappa}_m$

        \item For each interface: $\boldsymbol{w}_{nm}^n \leftarrow \boldsymbol{w}_{nm}$, $\boldsymbol{\kappa}_{nm}^{n} \leftarrow \boldsymbol{\kappa}_{nm}$
    \end{itemize}

\end{enumerate}
\end{tcolorbox}

\subsection{Energy considerations}\label{S:energy}

Consistency of mechanical work calculated at both scales must be assured  \cite{Perdahcoglu2011ConstitutiveHomogenization} leading to a Hill-Mandel condition \cite{Suquet1987ElementsMechanics}. 
For the present problem, this would require
\begin{equation}\label{eq:equality_work}
    \boldsymbol{\Sigma}:\boldsymbol{E}= \sum_{m \,\in\, L}  \phi_m \, \boldsymbol{\sigma}_m :\boldsymbol{\varepsilon}_m + \sum_{nm \,\in\, \mathcal{I}}  \, \boldsymbol{t}_{nm} \cdot \boldsymbol{w}_{nm} \,.
\end{equation}
We can use Equations (\ref{eq:compatibility2})--(\ref{eq:micro-balance3}) and (\ref{eq:Sigma}) to expand the right hand side as
\begin{align}
\sum_{m \,\in\, L}  \phi_m \, \boldsymbol{\sigma}_m :\boldsymbol{\varepsilon}_m + \sum_{nm \,\in\, \mathcal{I}}  \, \boldsymbol{t}_{nm} \cdot \boldsymbol{w}_{nm} 
&=   \sum_{m \,\in\, L}  \phi_m \, \boldsymbol{\sigma}_m :\left({\boldsymbol{E}} + {\boldsymbol{v}}_m \otimes^s \boldsymbol{n} \right) 
+ \boldsymbol{t} \cdot \left( \sum_{nm \,\in\, \mathcal{I}}  \, \boldsymbol{w}_{nm}\right) \notag \\
&= \boldsymbol{\Sigma}:\boldsymbol{E} + \sum_{m \,\in\, L}  \phi_m \, \left(\boldsymbol{\sigma}_m \cdot \boldsymbol{n} \right) \cdot {\boldsymbol{v}}_m + \boldsymbol{t} \cdot \left( \sum_{nm \,\in\, \mathcal{I}}  \, \boldsymbol{w}_{nm}\right) \notag \\
&= \boldsymbol{\Sigma}:\boldsymbol{E} + \boldsymbol{t} \cdot \left( \sum_{m \,\in\, L}  \phi_m \ {\boldsymbol{v}}_m +\sum_{nm \,\in\, \mathcal{I}}  \, \boldsymbol{w}_{nm} \right) \notag \\
&=  \boldsymbol{\Sigma}:\boldsymbol{E} \,,
\end{align}
confirming the Hill-Mandel condition is indeed satisfied.

\subsection{Recursive generalization}\label{S:multiscale}
\begin{figure}[t]
\centering
\includegraphics[width=4.0in]{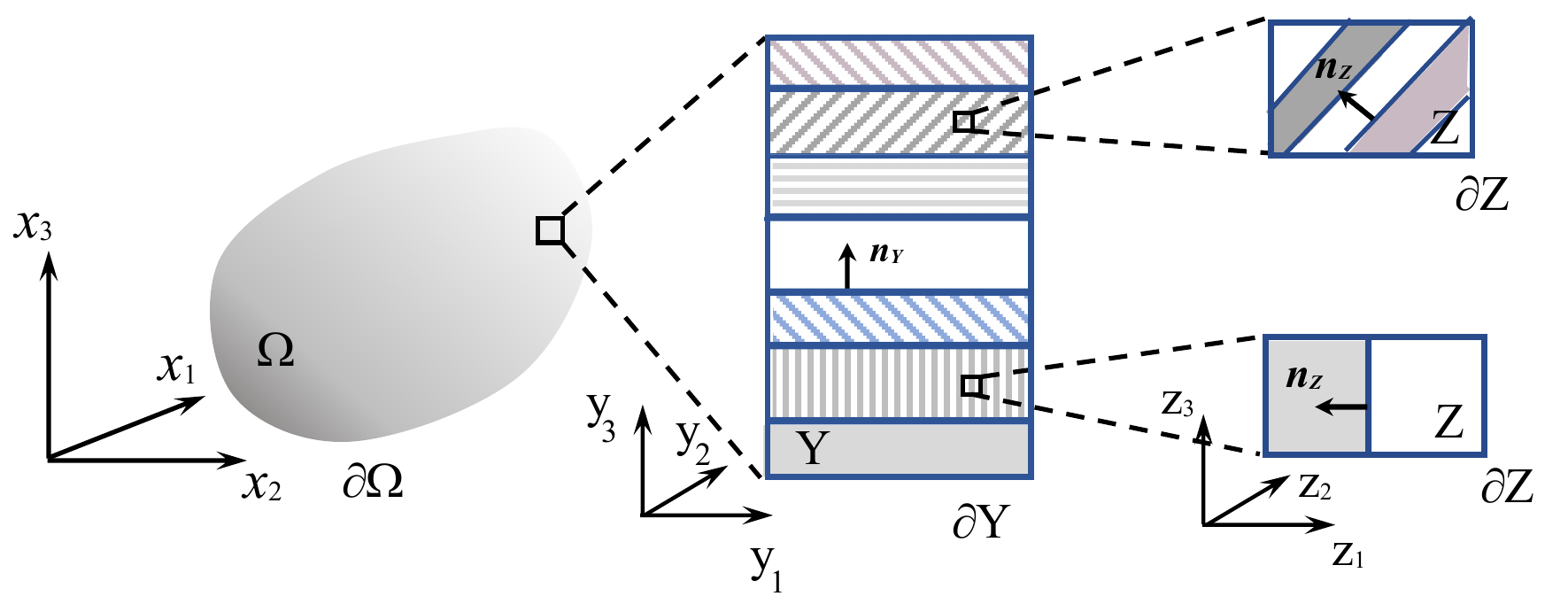}
\caption{Schematic representation of a three-scale domain. The macroscopic domain $\Omega$ is a periodic domain consisting of a multi-layer unit cell $Y$ with period $\epsilon_y$. Each layer is itself a periodic domain consisting of multi-layer unit cell $Z$ with period $\epsilon_z$.}
\label{fig:three-scale}
\end{figure}

 The formulation derived in the previous sections is based on the assumption that all layers are individually homogeneous. The constituent model for each layer, however, could itself arise from a homogenization process. For example, Figure \ref{fig:three-scale} shows a macroscopic domain $\Omega$ composed of a multi-layer unit cell $Y$ with period $\epsilon_y$ and layer normal $\boldsymbol{n}_y$. Each layer in unit cell $Y$, however, is itself a periodic domain unit cell $Z$ with period $\epsilon_z$ and layer normal $\boldsymbol{n}_z$.
 The two-scale approach presented in the previous sections can then be recursively extended to multiple scales, provided that all scales are sufficiently separate.  That is,
\begin{equation}
 \epsilon_z \ll \epsilon_y \ll1 \,.
\end{equation}
In this case, Step 2 in the numerical algorithm described in Algorithm 1 would consist of a recursive call to the homogenization routine to obtain the  response of each Z-periodic layer in $Y$. We note that a number of researchers have extended the asymptotic homogenization approach beyond two scales \cite{Bensoussan2011AsymptoticStructures}. See \cite{Ramirez-Torres2018ThreeTissues} for a review of multi-scale asymptotic homogenization.

\section{Numerical examples}\label{S:examples}

We now present several numerical examples using the proposed constitutive modeling framework.  The aim is to demonstrate its performance in capturing mechanical response of a variety of rock types, as observed from experiments.  This includes anisotropy in strength and elastic properties, as well as changing modes of failure. To demonstrate the framework's flexibility, we will incorporate a number of different micro-constitutive laws for the layers and interfaces. These sub-models are briefly described here, but more extensive discussions of their implementation and extensions can be found elsewhere \cite{Borja2013,White2014AnisotropicIntegration}.  More sophisticated formulations can be readily included as needed.

\subsection{Specific micro-constitutive models}
Each layer is assumed to be homogeneous, isotropic, and elastoplastic. The strain tensor is additively partitioned into elastic and plastic components as
\begin{equation}
\boldsymbol{\varepsilon} = \boldsymbol{\varepsilon}^e+\boldsymbol{\varepsilon}^p \,.
\end{equation}
The stress is computed from an isotropic, linear model as $\boldsymbol{\sigma} = \mathbb{C}^e:\boldsymbol{\varepsilon}^e$, with the elastic moduli $\mathbb{C}^e$ a function of the bulk modulus $K$ and Poisson ratio $\nu$.  Evolution of plastic strains in each layer is determined by a convex yield surface and an associative flow rule,
\begin{equation}\label{eq:yield_eps}
    f\left(\boldsymbol{\sigma}, \kappa\right) =0\,, \qquad     \dot{\boldsymbol{\varepsilon}}^{p}=\dot{\lambda} \frac{\partial f}{\partial \boldsymbol{\sigma}}\,,
\end{equation}
where $\kappa$ and $\dot{\lambda} \geq 0$ represent the hardening and plastic consistency parameters, respectively. The first yield function considered is the Drucker-Prager model,
\begin{equation}
    f\left(\boldsymbol{\sigma}_m, c\right)= q + \left( \tan \, \varphi \right) p - c
\end{equation}
in which $p=\operatorname{tr}\left(\boldsymbol{\sigma}\right)/3$ is the mean stress,  $\boldsymbol{s}=\boldsymbol{\sigma}-p\boldsymbol{1}$ is the deviatoric stress, and $q=\sqrt{3/2}\,\| \,\boldsymbol{s} \,\|$ is the von Mises stress. In addition, $\varphi$ and $c$ denote the friction angle and cohesion, respectively. Hardening (or softening) is introduced in the model as $\dot{c}=h \dot{\lambda}$ with modulus $h$ controlling the cohesion-hardening rate. 

The second yield criterion considered is the Modified Cam-Clay surface,
\begin{equation}
    f\left(\boldsymbol{\sigma}, p_c\right)= \frac{q^2}{M^2} + p\left(p-p_c\right)
\end{equation}
in which $p_c$ is the pre-consolidation pressure. The pre-consolidation pressure depends on volumetric plastic strains as $\dot{p_c} = h \, \operatorname{tr}(\dot{\boldsymbol{\varepsilon}}^p)$.

For each interface, the slip $\boldsymbol{w}$ is additively partitioned into elastic and inelastic components as
\begin{equation}
\boldsymbol{w} = \boldsymbol{w}^e + \boldsymbol{w}^p \,.
\end{equation}
The interface stiffness follows a linear model $\boldsymbol{t} = \mathbb{D}^e
\cdot \boldsymbol{w}^e$, with the elastic moduli $\mathbb{D}^e$ a function of the normal stiffness $k$ and shear stiffness $\mu$ of the interface.  Choosing very large stiffness coefficients will mimic an incompressible joint with rigid/plastic shear response. Inelastic slip is governed by a Coulomb yield surface and non-associative flow rule,
\begin{equation}
F(\boldsymbol{t},c) = t_s + (\tan \,\varphi)\, t_n - c \,, \qquad \dot{\boldsymbol{w}}^p = \dot \lambda \frac{\partial t_s}{\partial \boldsymbol{t}}\,.
\end{equation}
Here, $t_n = \boldsymbol{t} \cdot \boldsymbol{n}$ is the normal traction magnitude, $\boldsymbol{t}_s = \boldsymbol{t}-t_n \boldsymbol{n}$ is the shear traction, and $t_s=\|\boldsymbol{t}_s\|$ is the shear traction magnitude.  This model assumes a dilation-free inelastic slip.  See \cite{White2014AnisotropicIntegration} for additional details and extensions.

 \subsection{Synthetic layered rock experiments}
 
 \begin{table}[b]
\centering
\footnotesize
\begin{tabular} {lllrrr}
\toprule
 Parameter & Symbol &  Units &  Material A & Material B & Interface \\
 &&& (Drucker-Prager) & (Drucker-Prager) & (Coulomb)\\
\midrule
Phase fraction & $\phi_m$ & -- & 0.5 & 0.5 & --\\
Bulk modulus & K & MPa & 13,395 & 6,840 & --\\
Poisson's ratio & $\nu$ & -- & 0.23 & 0.21 & --\\
Normal stiffness & $k$ & MPa & -- & -- & $\infty$ \\
Shear stiffness & $\mu$ & MPa & -- & -- & $\infty$ \\
Friction angle & $\varphi$ & $^\circ$ &  35 & 18 & 18\\
Cohesion & c & MPa & 90 & 35 & 11\\
Hardening modulus & h & MPa & 0 & 0 & 0\\
\bottomrule
\end{tabular}
\caption{Parameters calibrated using data from the Tien et al. \cite{Tien2006AnRocks} experiments on synthetic layered rock.}
\end{table}

 \begin{figure}[p]
\centering
\includegraphics[width=3.in]{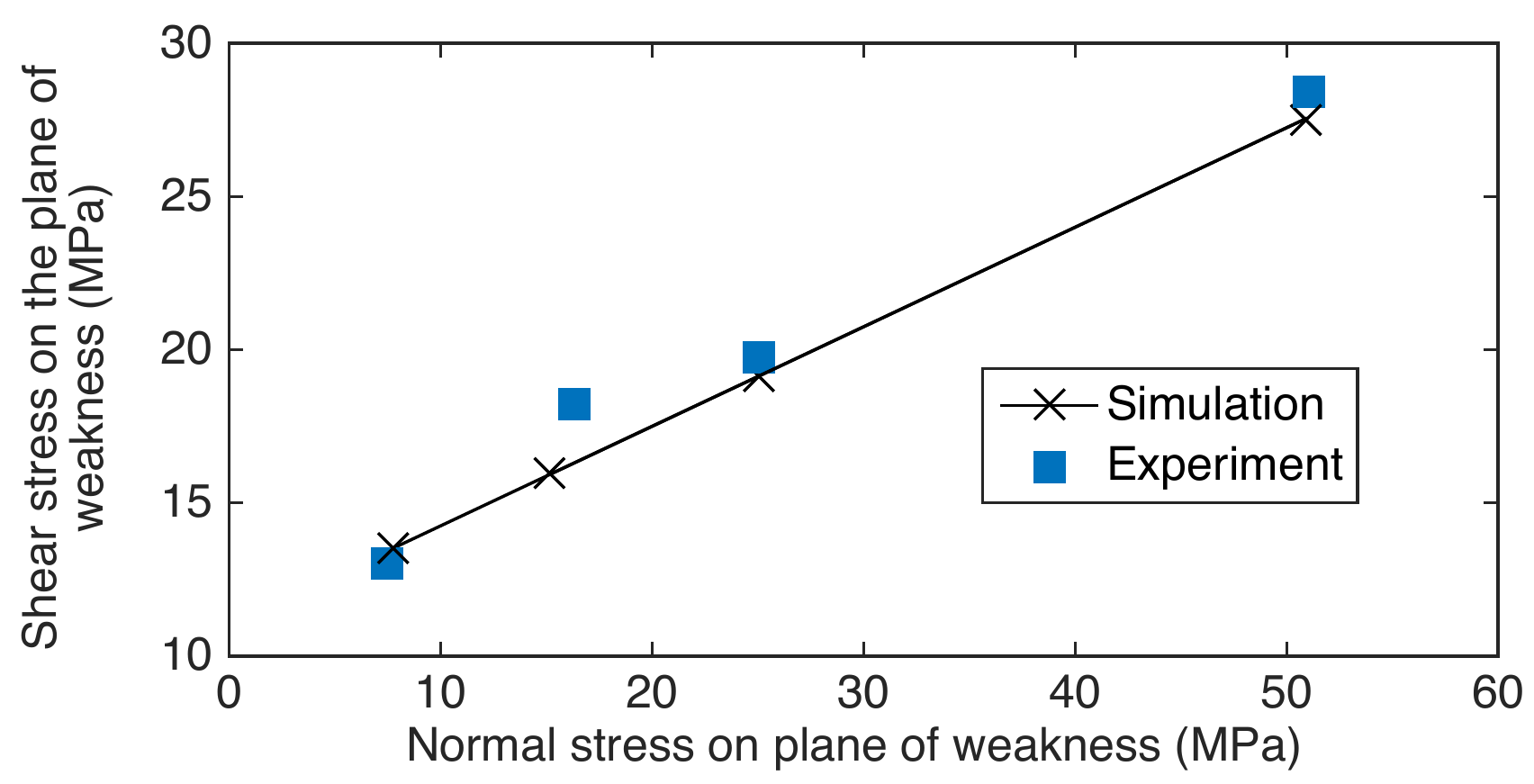}
\caption{Shear strength versus normal stress acting on an interface, for bedding plane orientation $\theta=60^{\circ}$, for the synthetic layered rock.}
\label{fig:shear_normal_tien}
\end{figure}

\begin{figure}[p]
\centering
\includegraphics[width=3.5in]{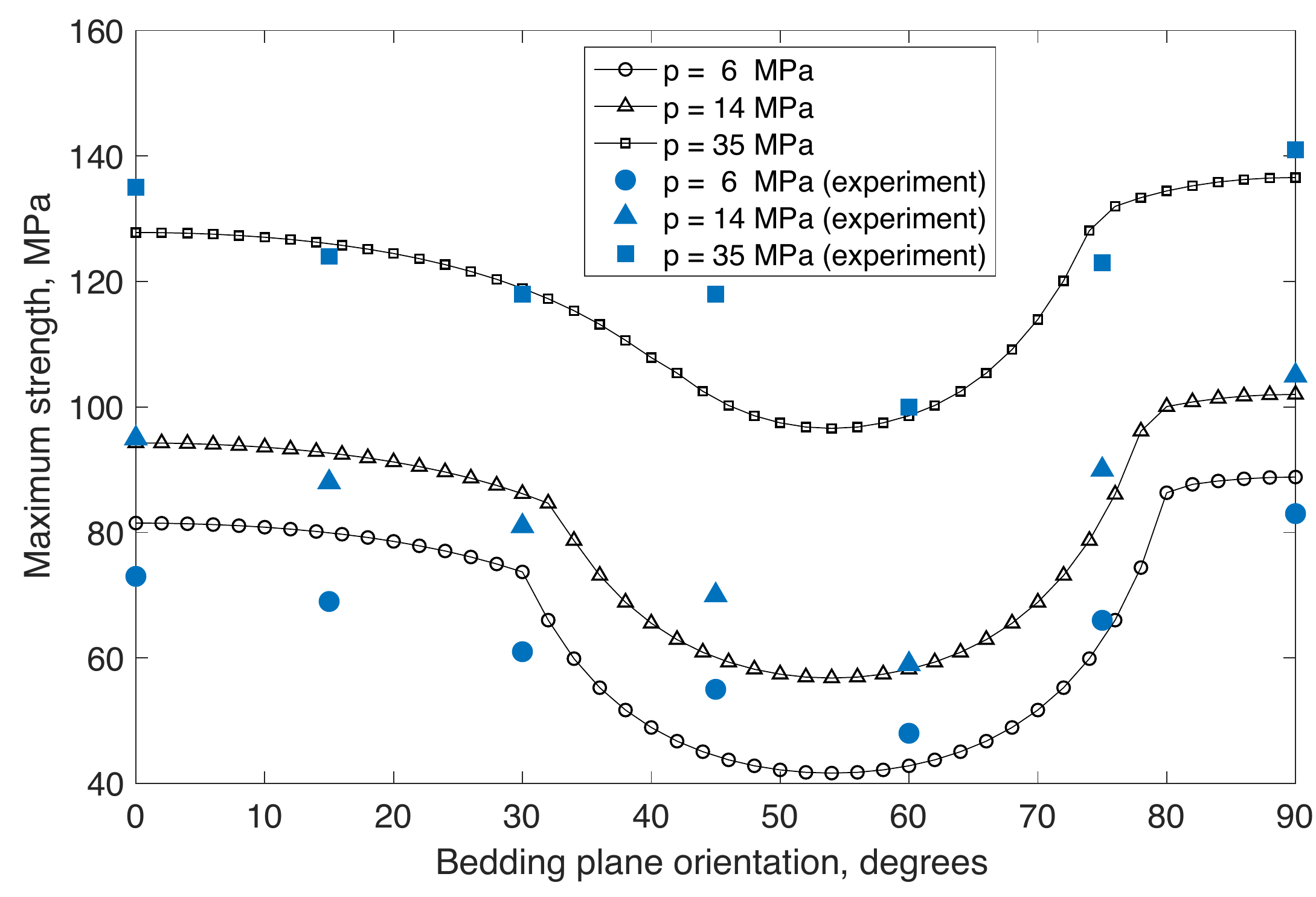}
\caption{Variation of maximum strength with respect to bedding plane orientation and confining pressure for the synthetic layered rock.}
\label{fig:strength_theta_tien}
\end{figure}

\begin{figure}[p]
\centering
\includegraphics[width=3.75in]{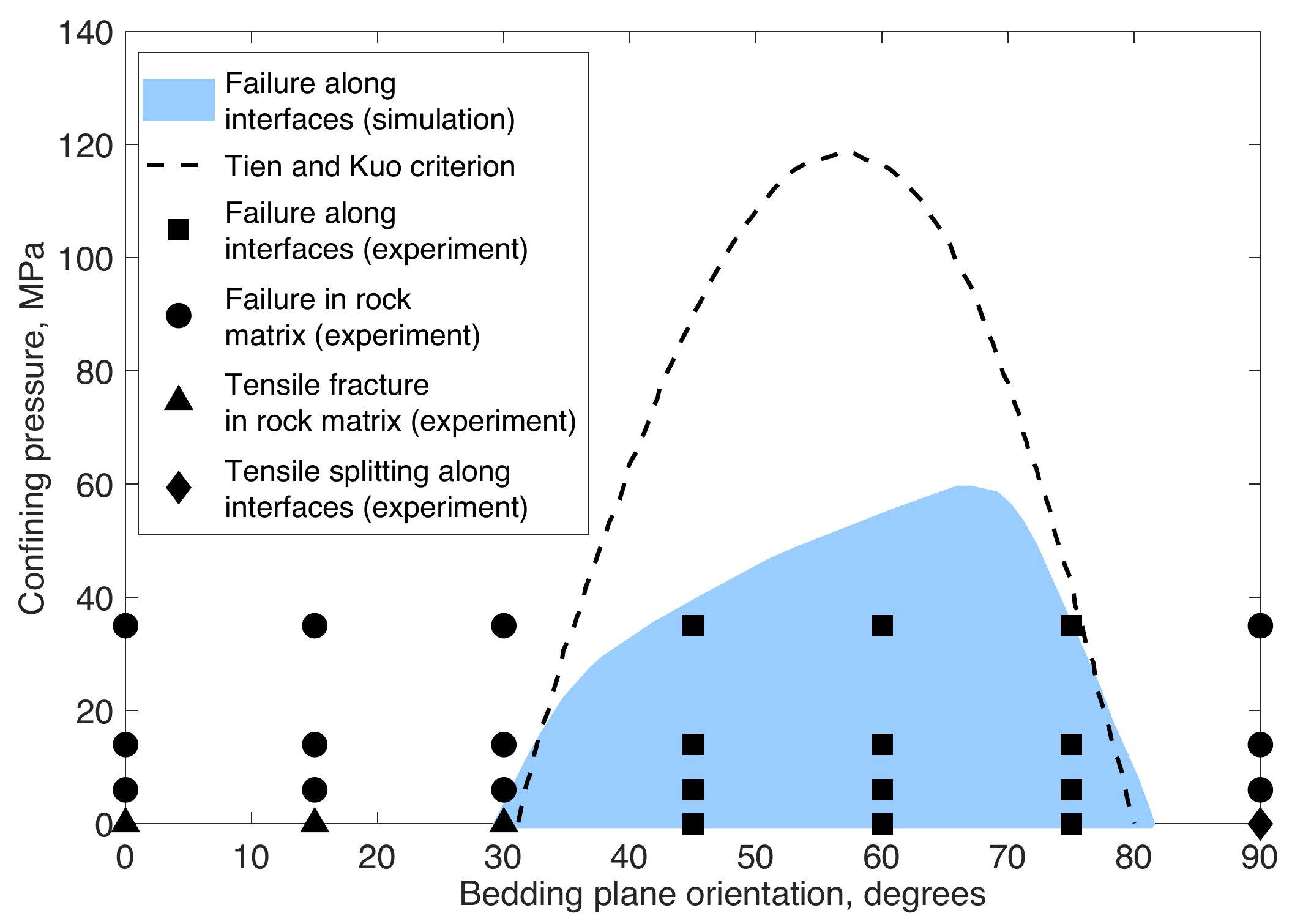}
\caption{Comparison of the failure modes as predicted by Tien and Kuo's criterion \cite{Tien2006AnRocks, Tien2001ARocks}, experiments, and simulations using the present method. The zone inside the marked regions corresponds to failure along discontinuities. }
\label{fig:failure_tien}
\end{figure}

   Tien et al. \cite{Tien2006AnRocks} used different weight ratios of cement, kaolinite, and water to produce two cementitious materials with different strength and stiffness values. They created a synthetic rock by layering these two materials in an alternating fashion. Subsequently, triaxial tests were conducted at various confining pressures and bedding plane orientations $\theta$, defined as the angle between the plane of the layers and the horizontal axis.  These experiments are an appealing test because the material microstructure is highly controlled, and multiple experiments in different orientations can be readily performed.

To mimic this synthetic microstructure, we use Drucker-Prager plasticity models for the two materials, and a rigid-plastic Coulomb model for the interface. Calibrated material parameters are summarized in Table 1.  Phase ratios for materials A and B are $\phi_A=\phi_B=0.5$. Since $\theta=60^{\circ}$ is reported as a case where failure occurs along the interface and not through the rock layers, we have used test results at this orientation to calibrate the Coulomb friction and cohesion parameters  (Figure \ref{fig:shear_normal_tien}). The intercept and slope on this plot correspond to the cohesion and tangent of friction angle, respectively.

 Figure \ref{fig:strength_theta_tien} compares peak strength results across a range of bedding orientations and confining pressures. It can be seen that the simple model presented here captures the complex strength anisotropy of the synthetic rock and its pressure dependence reasonably well. We note, however, that Tien et al.'s experimental results under zero confining pressure are not included in this figure, as tensile fractures in the rock matrix were observed during unconfined tests.  This particular failure mode is not possible using the Drucker-Prager model, though in the future a tensile failure criterion could be added to capture this particular mode if desired.

Different failure modes observed from the experiments are shown in Figure \ref{fig:failure_tien}. The regions corresponding to failure along an interface, as predicted by the proposed model and Tien and Kuo's criterion  \cite{Tien2006AnRocks, Tien2001ARocks} are also shown. It can be seen that the simulation results are consistent with the experimentally observed failure along the interfaces. At low confining pressures, the stress acting on the interface becomes critical and leads to failure for bedding plane orientations of $\sim$ $30^{\circ}$ to $80^{\circ}$.  At increased confining pressure, the strength of the interfaces increases due to the higher normal stress, and the range of bedding plane orientations at which the shear stress reaches a critical value decreases. At sufficiently high confining pressures, failure in the rock matrix will dominate.



 \subsection{Vaca Muerta shale}

We now explore triaxial test data on Vaca Muerta shale from Ambrose \cite{Ambrose2014FailureConditions}. Vaca Muerta shale has a very high organic content that is well-dispersed throughout the shale matrix. X-ray tomography images  obtained by Ambrose show laminations and weak planes in the samples, while on thin sections one dominant rock composition can be observed. Therefore, we propose to model the shale with a single matrix layer (phase fraction $\phi = 1.0$) with Drucker-Prager plasticity. Matrix layers are then separated by weak interfaces treated using a Coulomb frictional model, with elastic normal and shear compliance.

\begin{figure}[p]
\centering
\includegraphics[width=3.5in]{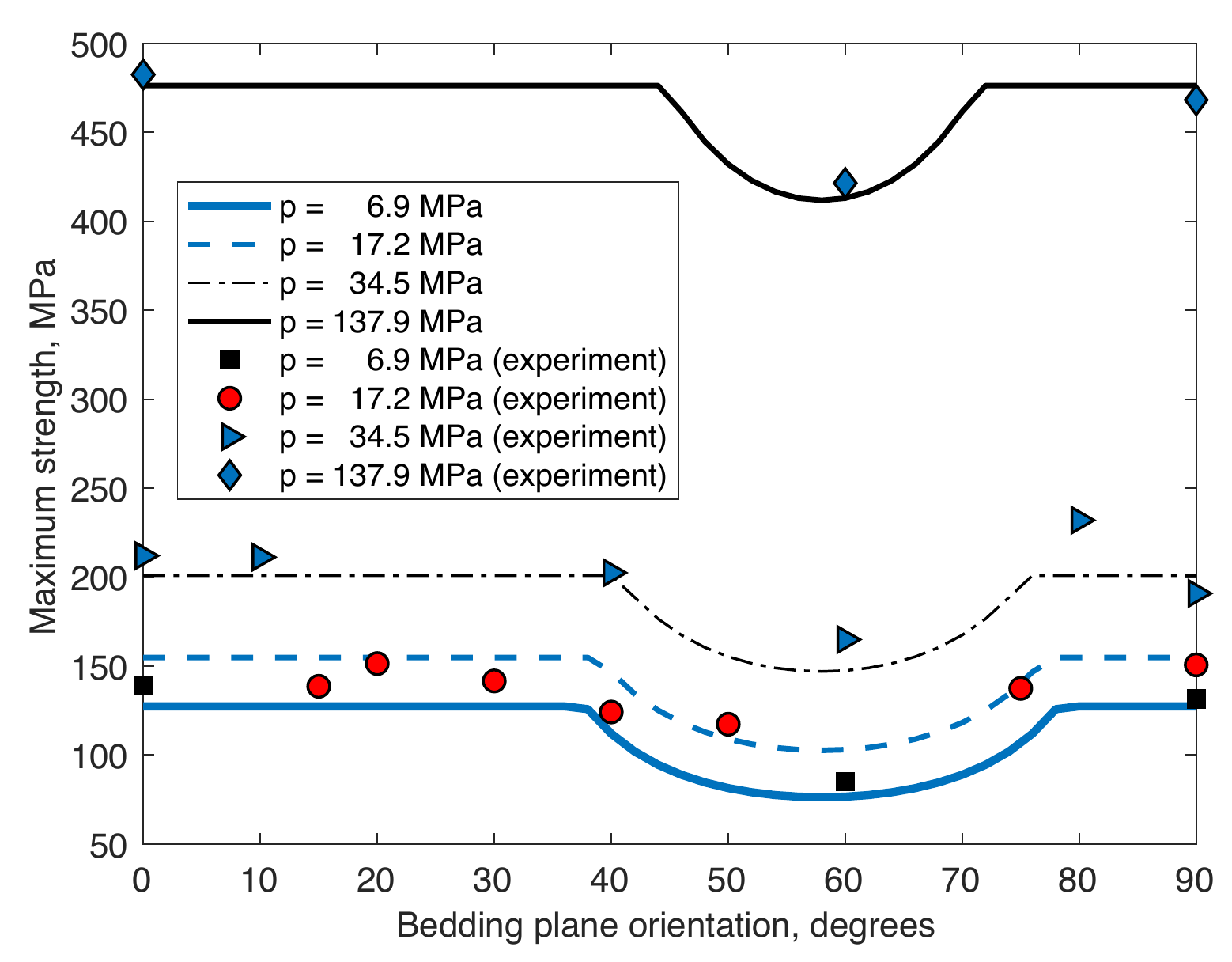}
\caption{Variation of maximum strength with respect to orientation of bedding planes obtained from triaxial tests on Vaca Muerta shale.}
\label{fig:strength_VM}
\end{figure}

\begin{figure}[p]
\centering
\includegraphics[width=3.5in]{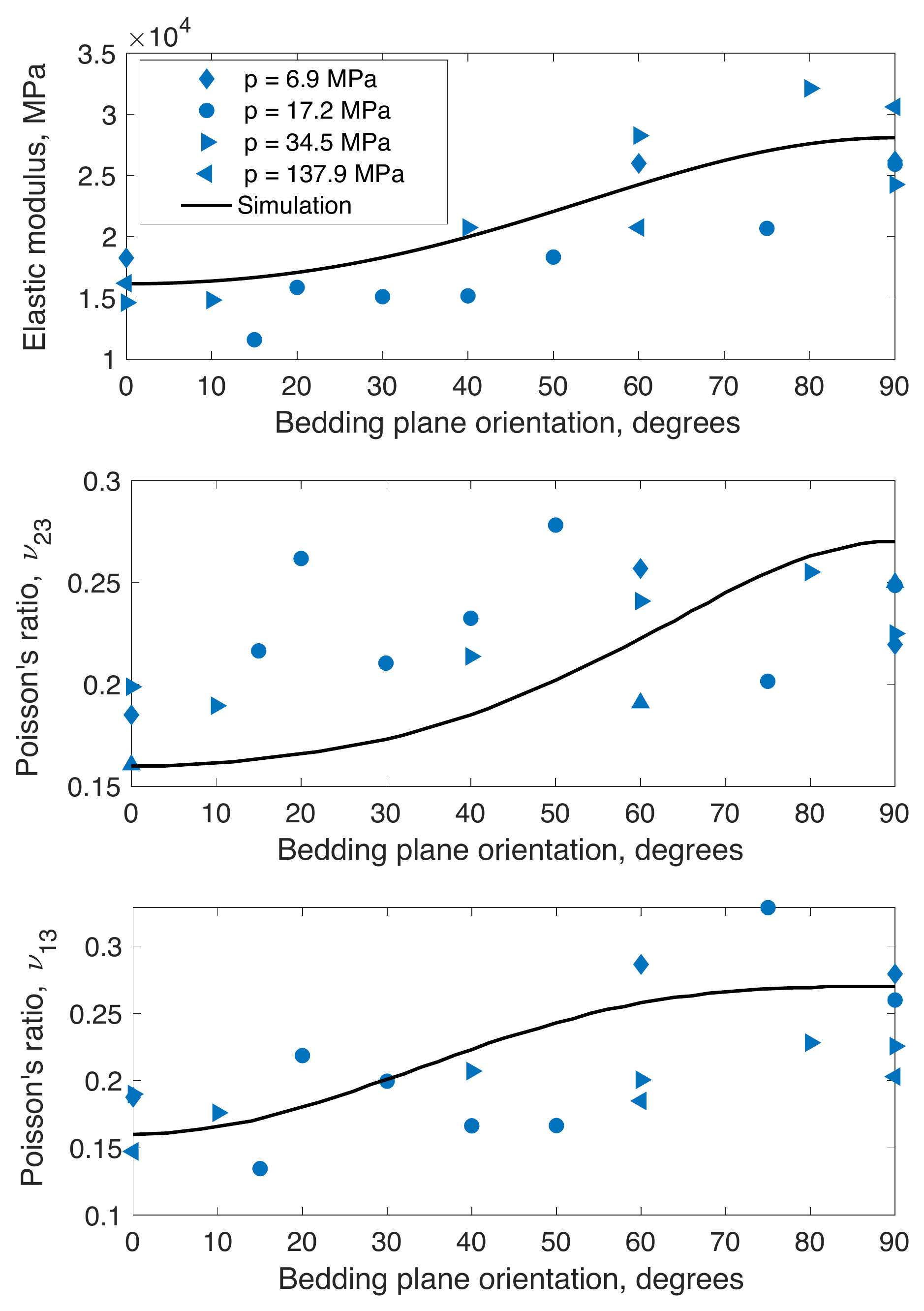}
\caption{Variation of macroscopic (apparent) elastic modulus in the loading direction (axis 3) and Poisson's ratios, with respect to orientation of bedding planes obtained from triaxial tests on Vaca Muerta shale.}
\label{fig:elasticity_VM}
\end{figure}

The calibrated model parameters are summarized in Table 2. Figure \ref{fig:strength_VM} shows the variation of maximum strength with respect to bedding plane orientations at confining pressures $p = 6.9, \, 17.2, \, 34.5, \, \text{and} \, 137.9$ MPa. The comparison illustrates that a simple micro-structural model is quite good at describing strength anisotropy in this material. Figure \ref{fig:elasticity_VM} also illustrates the resulting variation of macroscopic elastic modulus in the loading direction (axis 3) and the  two Poisson's ratios $\nu_{23}$ and $\nu_{13}$. Except for $\nu_{13}$, for which the experimental data appear very scattered, the other two elasticity parameters exhibit a smooth ascending trend as bedding plane orientations vary from $0^{\circ}$ to $90^{\circ}$, which is captured reasonably well by the present model.  We note that the selected elastic model has no intrinsic pressure dependence.  It is difficult to discern a consistent trend here given the experimental scatter in the data, however, so we have ignored this feature in the comparison.

   \begin{table}[h]
\centering
\footnotesize
\begin{tabular} {lllrr}
\toprule
 Parameter & Symbol &  Units &  Matrix & Interface \\
 &&& (Drucker-Prager) & (Coulomb)\\
\midrule
Bulk modulus & K & MPa & 17,390 & --\\
Poisson's ratio & $\nu$ & -- & 0.27 & --\\
Normal stiffness & $k$ & MPa & -- & 70,000 \\
Shear stiffness & $\mu$ & MPa & -- & 52,500 \\
Friction angle & $\varphi$ & $^\circ$ &  47 & 26\\
Cohesion & c & MPa & 70 & 18\\
Hardening modulus & h & MPa & 0 & 0\\
\bottomrule
\end{tabular}
\caption{Parameters calibrated using experimental data from Ambrose \cite{Ambrose2014FailureConditions}.}
\end{table}

 \subsection{Chichibu schist}
 
  \begin{table}[p]
\centering
\footnotesize
\begin{tabular} {lllrr}
\toprule
 Parameter & Symbol &  Units &  Matrix & Interface \\
 &&& (Drucker-Prager) & (Coulomb)\\
\midrule
Bulk modulus & K & MPa & 16880 & --\\
Poisson's ratio & $\nu$ & -- & 0.3 & --\\
Normal stiffness & $k$ & MPa & -- & $\infty$ \\
Shear stiffness & $\mu$ & MPa & -- & $\infty$ \\
Friction angle & $\varphi$ & $^\circ$ &  26.6 & 25.0\\
Cohesion & c & MPa & 300 & 32\\
Hardening modulus & h & MPa & 0 & 0\\
\bottomrule
\end{tabular}
\caption{Parameters calibrated using experimental data provided by Mogi \cite{MogiFLOWDruck}.}
\end{table}

 \begin{figure}[p]
\centering
\includegraphics[width=3.5in]{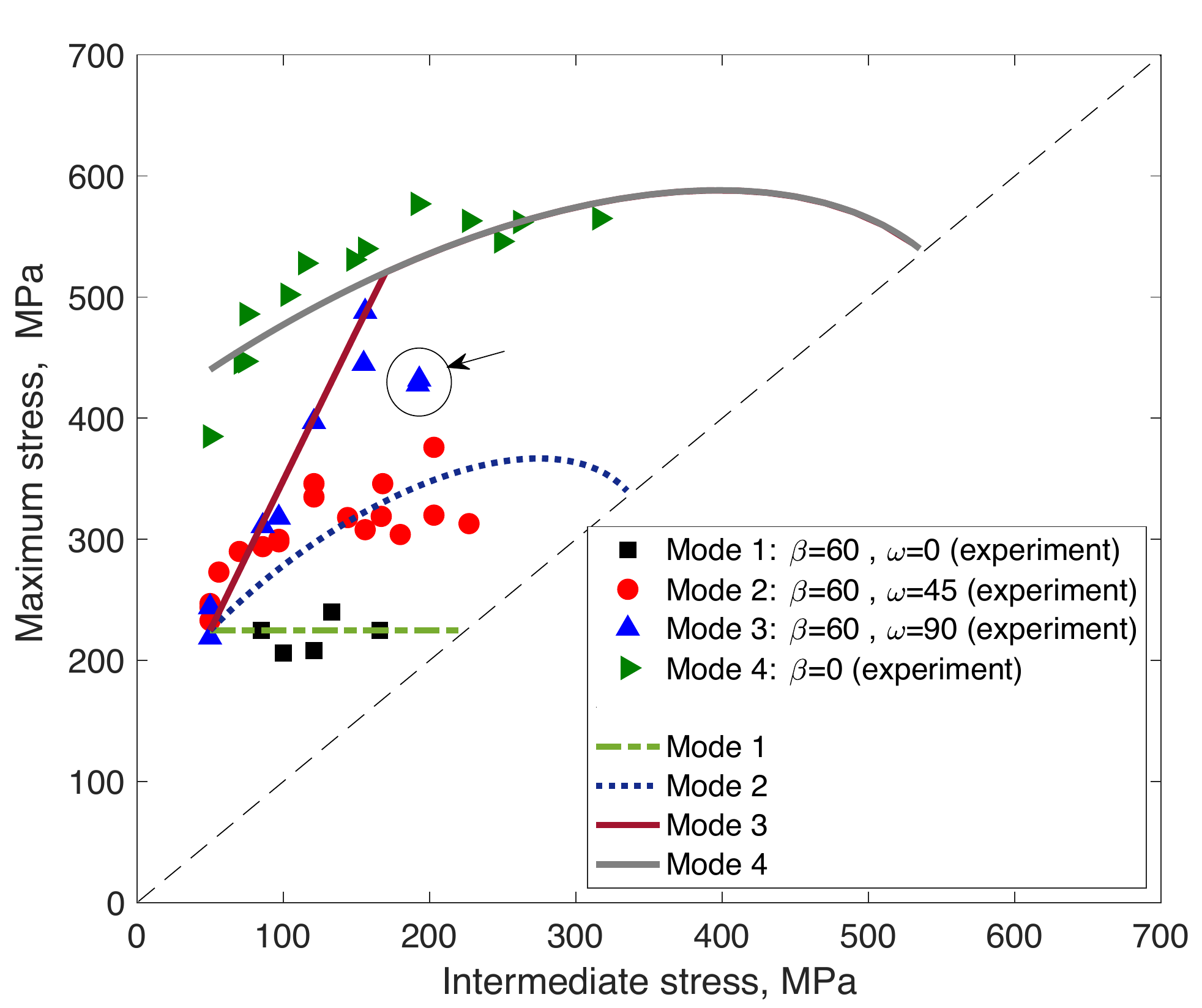}
\hspace{0.2in}
\includegraphics[width=1.75in]{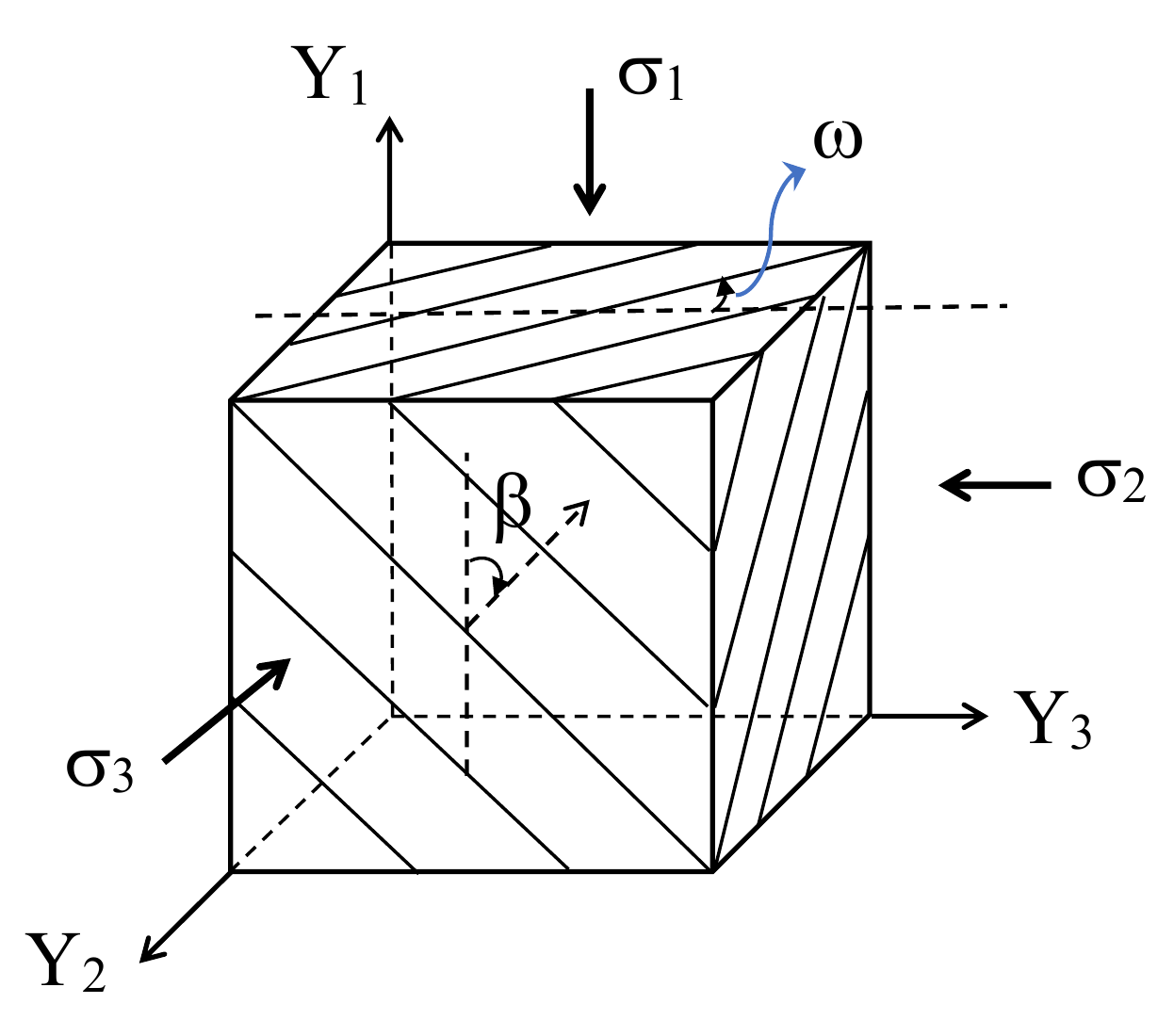}
\caption{Variation of maximum stress, $\sigma_1$, with respect to intermediate principal stress, $\sigma_2$. Minimum principal stress is $\sigma_3 = 50$ MPa for all cases.}
\label{fig:chichibu_strength}
\end{figure}

  \begin{figure}[p]
\centering
\includegraphics[width=3.5in]{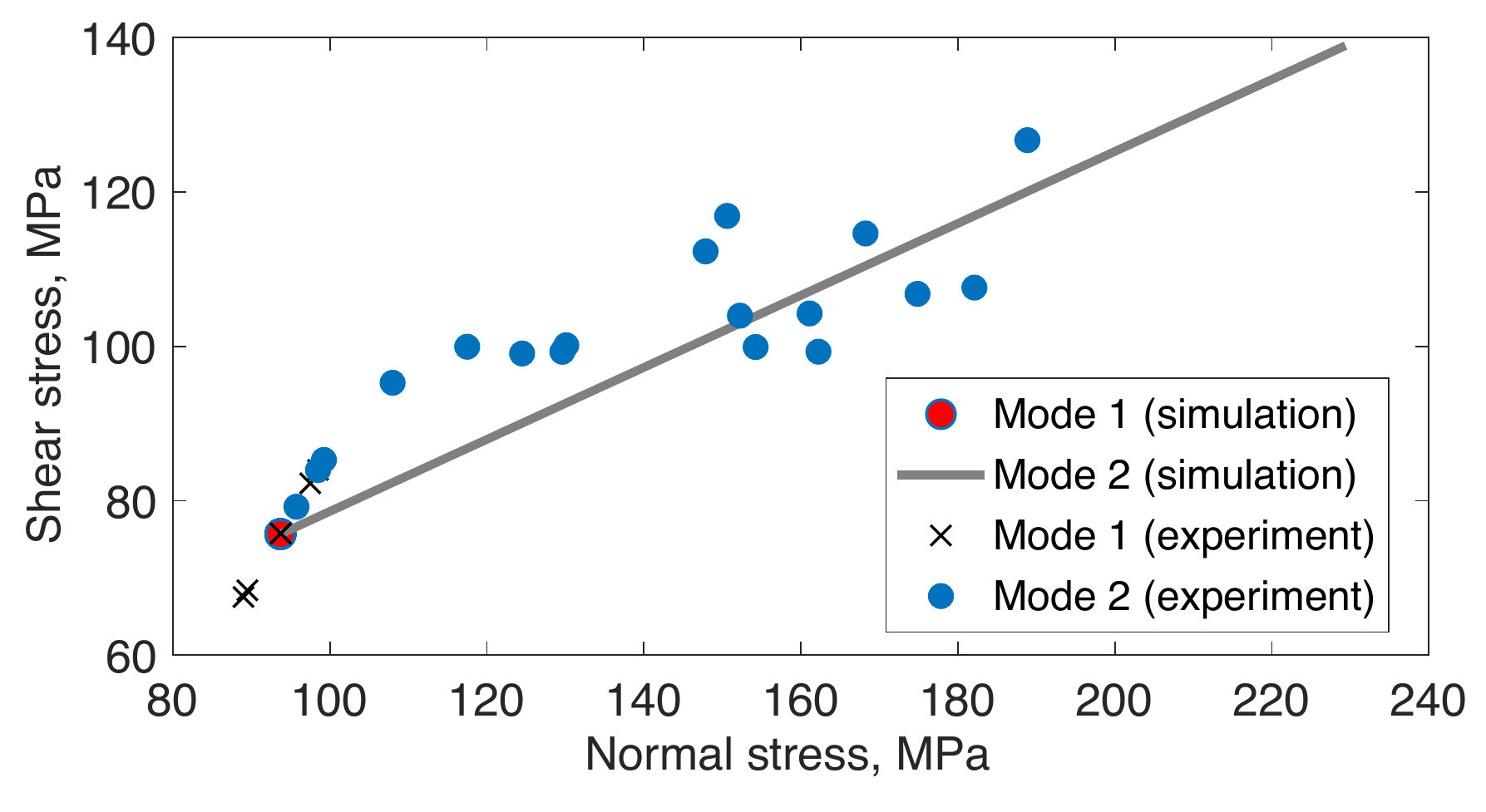}
\caption{Shear stress versus normal stress acting on the plane of weakness at failure, corresponding to Mode 1 and Mode 2.}
\label{fig:chichibu_shear_normal2}
\end{figure}

In this example, we use true triaxial test data from Mogi \cite{MogiFLOWDruck} for a schist rock from Chichibu on Honshu Island, Japan. This metamorphic crystalline rock has a densely foliated structure \cite{Kwasniewski2007ZACHOWANIEDYLATANCJA}. For modeling purposes, a single Drucker-Prager matrix type and rigid-plastic Coulomb interfaces are considered, with parameters provided in Table 3.

The orientation of the bedding planes in true triaxial tests is determined by two angles, $\beta$ and $\omega$, which are defined with respect to the principal stress directions $\sigma_1>\sigma_2>\sigma_3$. Figure \ref{fig:chichibu_strength} shows variations of maximum strength in the direction of $\sigma_1$ versus intermediate stress $\sigma_2$, while $\sigma_3=50$ MPa is held constant in all experiments. The results obtained from simulations are compared with experiments in four different modes, showing good agreement across a variety of failure mechanisms. Mode 1 ($\beta=60^{\circ}$, $\omega=0^{\circ}$) and Mode 2 ($\beta=60^{\circ}$, $\omega=45^{\circ}$) lead to failure along the interfaces, while in Mode 4 ($\beta=0^{\circ}$) failure is only observed in the rock matrix. Mode 3  ($\beta=60^{\circ}$, $\omega=90^{\circ}$) exhibits a transition from failure along interfaces at smaller values of $\sigma_2$ to failure in the rock matrix at higher values of $\sigma_2$. Note that the data points marked with an arrow in Figure \ref{fig:chichibu_strength} correspond to a transitional mode of failure and are disregarded in the calibration process. Figure \ref{fig:chichibu_shear_normal2} shows the variation of shear stress versus normal stress acting on the interfaces at failure for Modes 1 and 2. For Mode 1, we have $\omega=0$.  Changes in the intermediate stress $\sigma_2$ in the simulations do not affect the stress conditions on the interface, and we obtain only one point corresponding to this mode (red circle in Figure \ref{fig:chichibu_shear_normal2}). Experimental data corresponding to Mode 1 shows some small variation in the maximum strength with respect to $\sigma_2$, while these changes are much more pronounced for other modes with non-zero angle $\omega$.

\subsection{Brittle-ductile material}

In this final example, we focus on capturing a transition from brittle to ductile behavior in geologic materials.  This example is motivated by oil shale experiments \cite{white2017thermoplasticity} which show a clear transition from brittle to ductile behavior with increasing organic content. For this purpose, we present results for a hypothetical material with two perfectly bonded layers (no weak interface). Material A has a ductile behavior modeled with Modified Cam-Clay. Layer B has a brittle behavior modeled with Drucker-Prager. Material parameters are indicated in Table 4. We will vary the phase fraction of each material to control the fraction of brittle versus ductile constituents.

\begin{figure}[t]
\centering
\begin{subfigure}[t]{0.75\textwidth}
\centering
\includegraphics[width=3.5in]{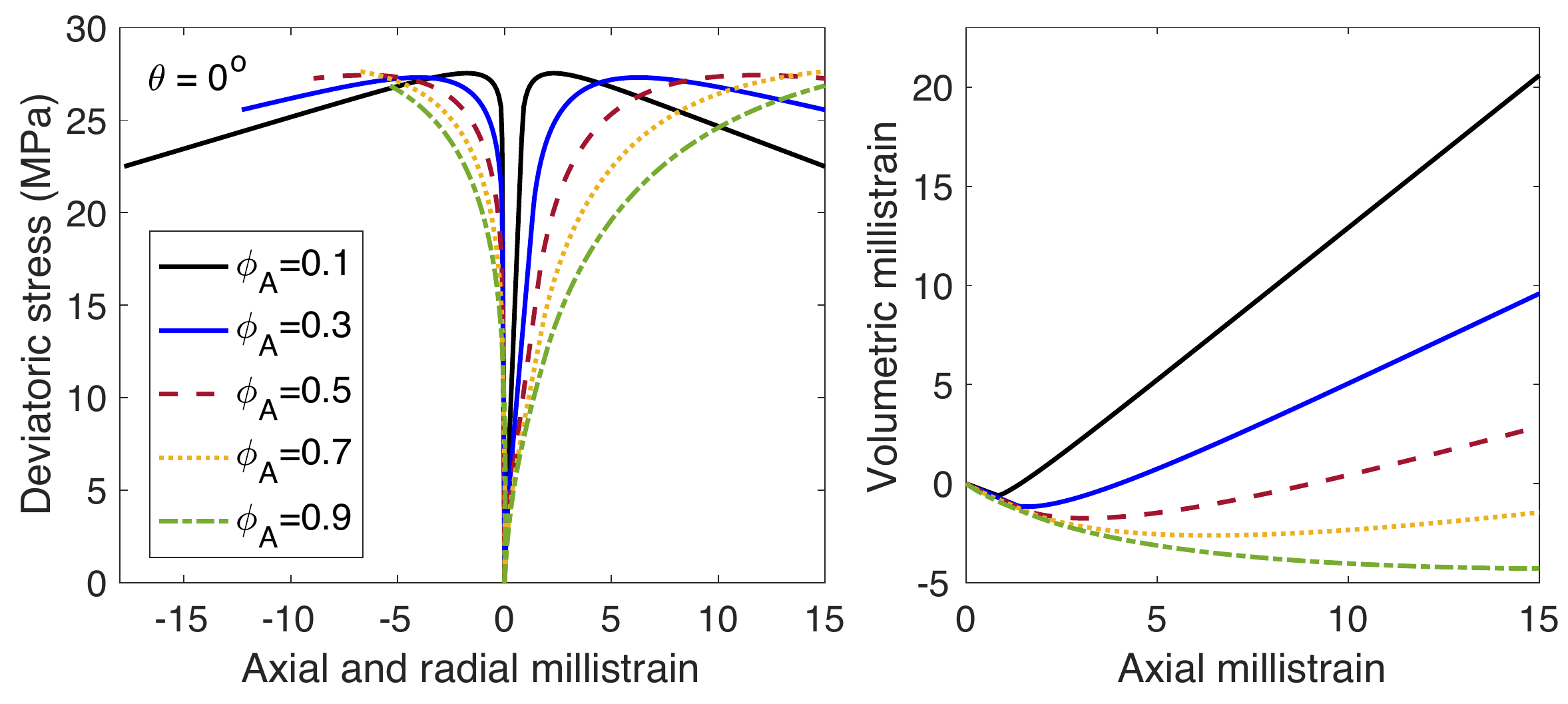}
\caption*{}
\label{fig:varying_phi_theta_0}
\end{subfigure}
\begin{subfigure}[H]{0.75\textwidth}
\centering
\includegraphics[width=3.5in]{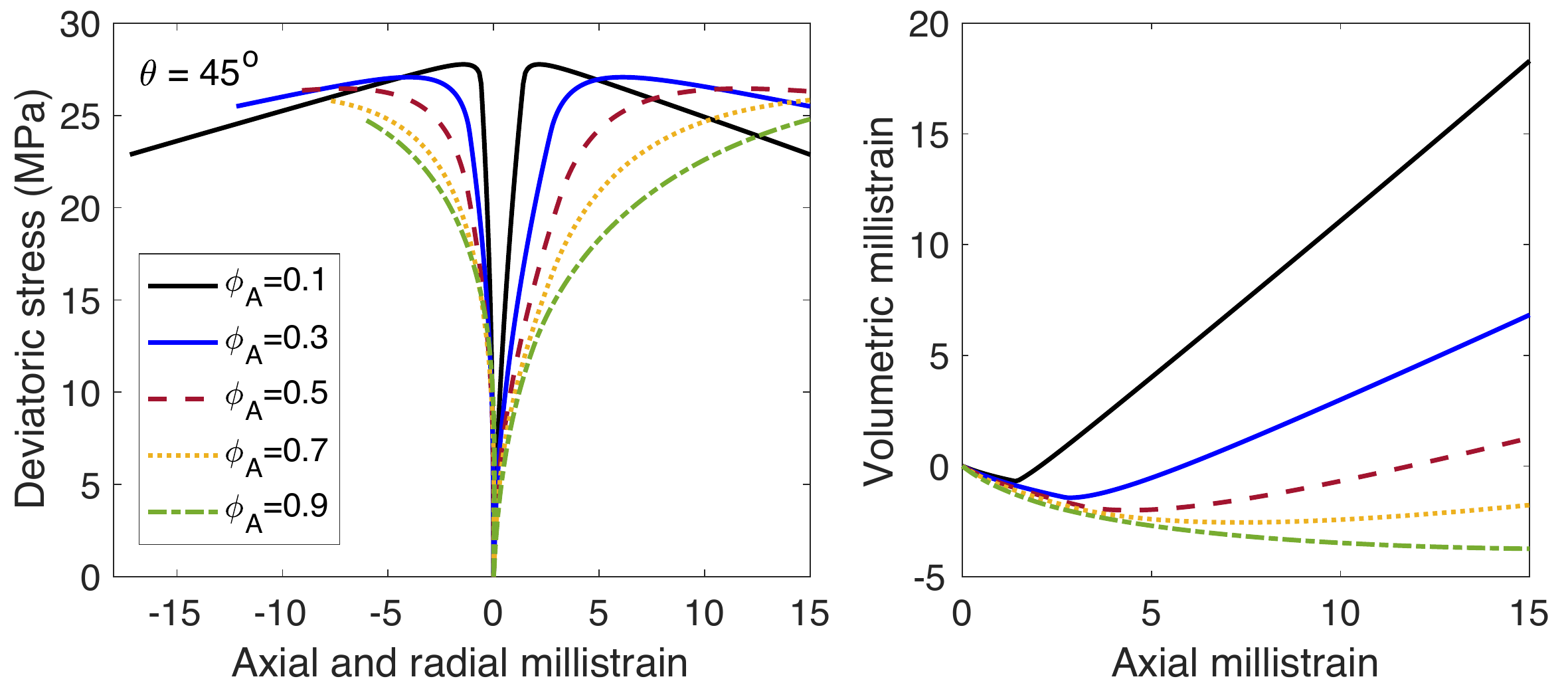}
\caption*{}
\label{fig:varying_phi_theta_45}
\end{subfigure}
\begin{subfigure}[H]{0.75\textwidth}
\centering
\includegraphics[width=3.5in]{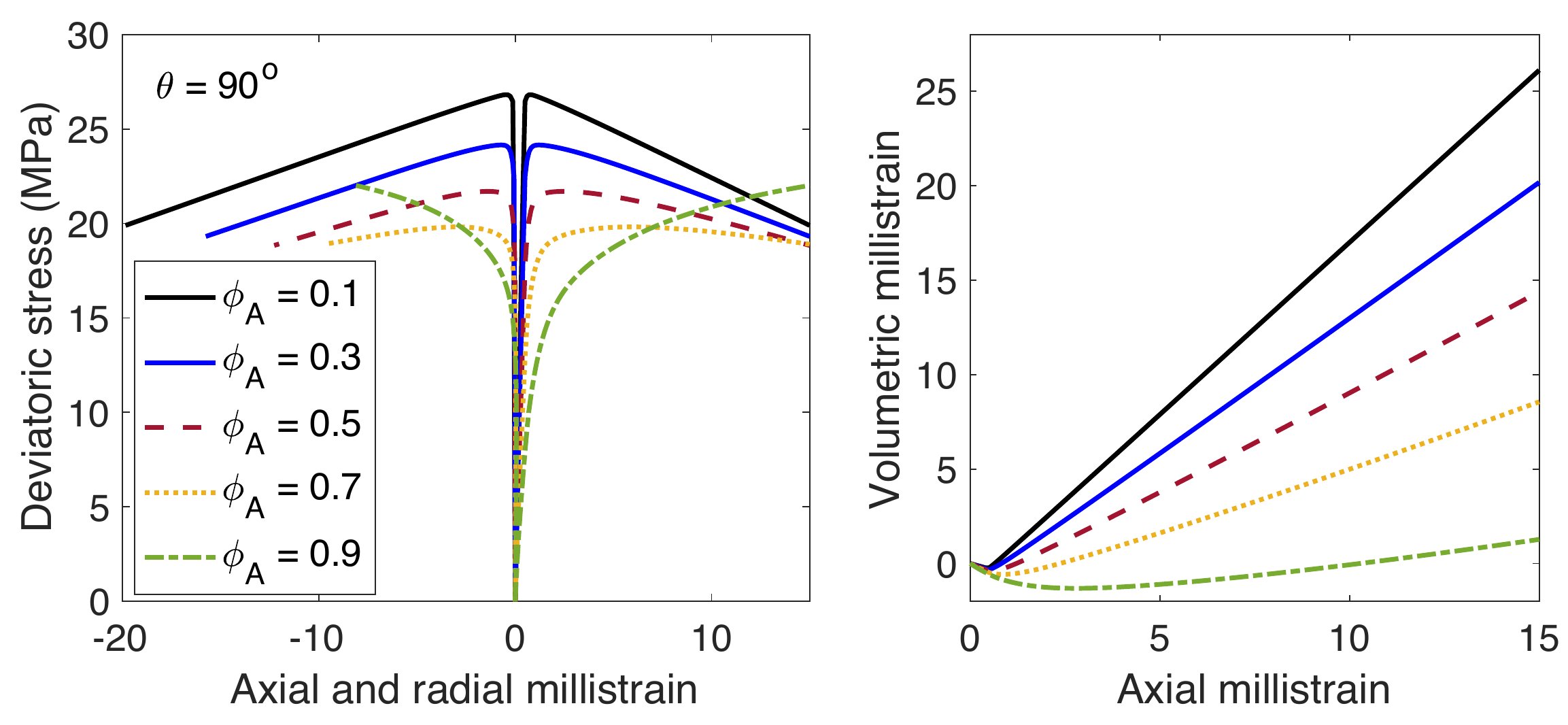}
\caption*{}
\label{fig:varying_phi_theta_90}
\end{subfigure}
\caption{Variation of deviatoric stress versus axial and radial strains (left) and volumetric strain versus axial strain (right) for phase ratios of the ductile layer varying between $0.1$ to $0.9$.  Panels from (top) to (bottom) indicate bedding plane orienations of $\theta=0^{\circ},45^{\circ}, \text{and }90^{\circ}$.}
\label{fig:varying-phi}
\end{figure}

Triaxial simulations are conducted with a constant confining pressure of $p=10$ MPa and bedding plane orientations of $\theta=0^{\circ}, \, 45^{\circ}, \, \text{and } 90^{\circ}$. Figure \ref{fig:varying-phi} illustrates variation of the macroscopic deviatoric stress versus macroscopic axial and radial strains, as well as variation of macroscopic volumetric strain versus axial strain for various ratios of phase $A$, from $\phi_A=0.1$ to $\phi_A=0.9$. It should be noted that the two components of radial strains in two different directions are not identical, and only one is shown here for clarity. As expected, upon increasing the ratio of the ductile phase, the material shows a more ductile overall macroscopic behavior. We also observe a significant reduction in compaction of the material for $\theta=45^{\circ}$ and $90^{\circ}$ upon increasing the ratio of the ductile phase. A transition from macroscopic compaction to dilation is observed for $\theta=0^{\circ}$, when the ductile phase comprises around $70\%$ of the volume.

\begin{table}[h]
\centering
\footnotesize
\begin{tabular} {lllrr}
\toprule
 Parameter & Symbol &  Units &  Material A & Material B \\
 &&& (Modified Cam-Clay) & (Drucker-Prager)\\
\midrule
Bulk modulus & K & MPa & 26.7 & 40\\
Poisson's ratio & $\nu$ & -- & 0.25 & 0.25\\
CSL slope & $M$ & -- & 1.5 & --\\
Pre-consolidation pressure & $p_c$ & MPa & 10 & --\\
Friction angle & $\varphi$ & $^\circ$ &  -- & 50\\
Cohesion & c & MPa & -- & 5\\
Hardening modulus & h & MPa & 5000 & -200\\
\bottomrule
\end{tabular}
\caption{Parameters for the brittle-ductile material example.}
\end{table}

\section{Concluding remarks}
We have presented an inelastic homogenization framework for layered materials with planes of weakness. Key features of the proposed approach include explicit representation of layered microstructures,  potential for slip along interface discontinuities, and the ability to include arbitrary combinations of micro-constituent models within the same numerical framework. This approach was applied to a diverse variety of experimental data---e.g. layered synthetic rocks, Vaca Muerta shale, and Chichibu schist. The proposed model performs well in capturing both strength and stiffness anisotropy. Moreover, it was shown that the models can describe failure modes in the rock matrix and along planes of weakness in a manner consistent with experimental observations. Numerical examples for a hypothetical bi-layer material also demonstrate the ability to describe complex phenomenology for materials with mixed brittle and ductile constituents.  

 A number of failure criteria for transversely isotropic rocks have been developed in the literature \cite{Jaeger1960ShearRocks,Hoek1990EstimatingCriterion,Hoek2002Hoek-BrownEdition,Tien2001ARocks}. A useful example is Jaeger's plane of weakness model \cite{Jaeger1960ShearRocks}, which employs a Mohr-Coulomb failure criterion for the rock matrix and sliding on planes of weakness. The present method can exactly mimic Jaeger's model when a unit cell with one rock layer and one rigid-plastic interface is used.  This model, however, will describe the complete stress-strain response of the material, not just the failure envelope.
 
Anisotropic elasto-plastic continuum models have also been developed in the literature to capture anisotropy in the mechanical response of transversely isotropic materials---e.g. by extension of isotropic Drucker-Prager \cite{Pariseau1968PlasticitySoil} and Cam-Clay type models \cite{Semnani2016ThermoplasticityPlasticity, Nova1986AnRocks}. These models have also been applied to strength anisotropy of rocks \cite{zhao2018,Ambrose2014FailureConditions, Pariseau1968PlasticitySoil}. One advantage of the present framework is that the overall anisotropic behavior emerges naturally from the interaction of isotropic layers within a clearly defined microstructure. We therefore view the proposed homogenization methodology as a natural framework for designing anisotropic material models, allowing model parameters to be directly connected to a microstructural model in a physically motivated way.

\section*{Acknowledgements}
Funding for this work was provided by Total S.A. through the FC-MAELSTROM Project. Portions of this work were performed under the auspices of the U.S. Department of Energy by Lawrence Livermore National Laboratory under Contract DE-AC52-07-NA27344.

\section*{References}
\bibliographystyle{abbrv}
\bibliography{Mendeley.bib}

\begin{thebibliography}{100}
\expandafter\ifx\csname url\endcsname\relax
  \def\url#1{\texttt{#1}}\fi
\expandafter\ifx\csname urlprefix\endcsname\relax\def\urlprefix{URL }\fi
\expandafter\ifx\csname href\endcsname\relax
  \def\href#1#2{#2} \def\path#1{#1}\fi

\bibitem{Salamon1968ElasticMass}
M.~D. Salamon, {Elastic moduli of a stratified rock mass}, International
  Journal of Rock Mechanics and Mining Sciences \& Geomechanics Abstracts 5~(6)
  (1968) 519--527.
\newblock \href {http://dx.doi.org/10.1016/0148-9062(68)90039-9}
  {\path{doi:10.1016/0148-9062(68)90039-9}}.

\bibitem{Sawicki1978OnSoil}
A.~Sawicki, {On application of effective moduli theory to layered soil},
  Rozprawy Hydrotechniczne 39 (1978) 3--13.

\bibitem{Semnani2016ThermoplasticityPlasticity}
S.~J. Semnani, J.~A. White, R.~I. Borja, {Thermoplasticity and strain
  localization in transversely isotropic materials based on anisotropic
  critical state plasticity}, International Journal for Numerical and
  Analytical Methods in Geomechanics 40~(18) (2016) 2423--2449.
\newblock \href {http://dx.doi.org/10.1002/nag.2536}
  {\path{doi:10.1002/nag.2536}}.

\bibitem{Pariseau1968PlasticitySoil}
W.~G. Pariseau, {Plasticity Theory For Anisotropic Rocks And Soil}, in: The
  10th US Symposium on Rock Mechanics (USRMS), 1968.

\bibitem{Tien2001ARocks}
Y.~M. Tien, M.~C. Kuo, {A failure criterion for transversely isotropic rocks},
  International Journal of Rock Mechanics and Mining Sciences 38~(3) (2001)
  399--412.

\bibitem{Drzal1990TheProperties}
L.~T. Drzal, {The role of the fiber-matrix interphase on composite properties},
  Vacuum 41~(7-9) (1990) 1615--1618.
\newblock \href {http://dx.doi.org/10.1016/0042-207X(90)94034-N}
  {\path{doi:10.1016/0042-207X(90)94034-N}}.

\bibitem{Jaeger1960ShearRocks}
J.~C. Jaeger, {Shear failure of anistropic rocks}, Geological Magazine 97~(1)
  (1960) 65--72.
\newblock \href {http://dx.doi.org/10.1017/S0016756800061100}
  {\path{doi:10.1017/S0016756800061100}}.

\bibitem{Hoek1990EstimatingCriterion}
E.~Hoek, {Estimating Mohr-Coulomb friction and cohesion values from the
  Hoek-Brown failure criterion}, International Journal of Rock Mechanics and
  Mining Sciences {\&} Geomechanics Abstracts 27~(3) (1990) 227--229.

\bibitem{Hoek2002Hoek-BrownEdition}
E.~Hoek, C.~Carranza-Torres, B.~Corkum, {Hoek-Brown failure criterion-2002
  edition}, Proceedings of NARMS-Tac 1 (2002) 267--273.

\bibitem{Kanoute2009MultiscaleReview}
P.~Kanout{\'{e}}, D.~P. Boso, J.~L. Chaboche, B.~A. Schrefler, {Multiscale
  methods for composites: A review}, Archives of Computational Methods in
  Engineering 16~(1) (2009) 31--75.
\newblock \href {http://dx.doi.org/10.1007/s11831-008-9028-8}
  {\path{doi:10.1007/s11831-008-9028-8}}.

\bibitem{Geers2017HomogenizationProblems}
M.~G.~D. Geers, V.~G. Kouznetsova, K.~Matou{\v{s}}, J.~Yvonnet, {Homogenization
  Methods and Multiscale Modeling: Nonlinear Problems}, in: E.~Stein,
  R.~de~Borst, T.~J. Hughes (Eds.), Encyclopedia of Computational Mechanics
  Second Edition, John Wiley {\&} Sons, Ltd., 2017.
\newblock \href {http://dx.doi.org/10.1002/9781119176817.ecm2107}
  {\path{doi:10.1002/9781119176817.ecm2107}}.

\bibitem{Pavliotis2008Multi-scaleHomogenization}
G.~A. Pavliotis, A.~M. Stuart, {Multi-scale methods: Averaging and
  homogenization}, Springer Science {\&} Business Media, New York, 2008.

\bibitem{OrtolanoGonzalez2013AMaterials}
J.~M. Ortolano~Gonz{\'{a}}lez, J.~A. Hern{\'{a}}ndez~Ortega,
  X.~Oliver~Olivella, {A comparative study on homogenization strategies for
  multi-scale analysis of materials}, Centre Internacional de M{\`{e}}todes
  Num{\`{e}}rics en Enginyeria (CIMNE), 2013.

\bibitem{Vondrejc2013FFT-basedProject}
J.~Vondrejc,
  \href{https://www.researchgate.net/publication/292146929}{{FFT-based method
  for homogenization of periodic media: Theory and applications FFT-based
  micro-mechanical solver View project}}, Ph.D. thesis, Czech Technical
  University in Prague (2013).
\newblock \href {http://dx.doi.org/10.13140/RG.2.1.2534.2489}
  {\path{doi:10.13140/RG.2.1.2534.2489}}.
\newline\urlprefix\url{https://www.researchgate.net/publication/292146929}

\bibitem{Moulinec1994AComposites}
H.~Moulinec, P.~Suquet, {A fast numerical method for computing the linear and
  nonlinear mechanical properties of composites}, C. R Acad. Sci. Paris 318
  (1994) 1417--1423.

\bibitem{Michel1999EffectiveApproach}
J.~C. Michel, H.~Moulinec, P.~Suquet, {Effective properties of composite
  materials with periodic microstructure: A computational approach}, Computer
  Methods in Applied Mechanics and Engineering 172~(1-4) (1999) 109--143.
\newblock \href {http://dx.doi.org/10.1016/S0045-7825(98)00227-8}
  {\path{doi:10.1016/S0045-7825(98)00227-8}}.

\bibitem{Dvorak1992TransformationMaterials}
G.~J. Dvorak, \href{http://rspa.royalsocietypublishing.org/}{{Transformation
  field analysis of inelastic composite materials}}, Proc. R. Soc. Lond. A 437
  (1992) 311--327.
\newline\urlprefix\url{http://rspa.royalsocietypublishing.org/}

\bibitem{Marfia2016ComputationalPhenomena}
S.~Marfia, E.~Sacco, {Computational homogenization of composites experiencing
  plasticity, cracking and debonding phenomena}, Computer Methods in Applied
  Mechanics and Engineering 304 (2016) 319--341.
\newblock \href {http://dx.doi.org/10.1016/j.cma.2016.02.007}
  {\path{doi:10.1016/j.cma.2016.02.007}}.

\bibitem{Sepe2013AField}
V.~Sepe, S.~Marfia, E.~Sacco, {A nonuniform TFA homogenization technique based
  on piecewise interpolation functions of the inelastic field}, International
  Journal of Solids and Structures 50~(5) (2013) 725--742.
\newblock \href {http://dx.doi.org/10.1016/j.ijsolstr.2012.11.005}
  {\path{doi:10.1016/j.ijsolstr.2012.11.005}}.

\bibitem{Michel2003NonuniformAnalysis}
J.~C. Michel, P.~Suquet, {Nonuniform transformation field analysis},
  International journal of solids and structures 40~(25) (2003) 6937--6955.
\newblock \href {http://dx.doi.org/10.1016/S0020-7683(03)00346-9}
  {\path{doi:10.1016/S0020-7683(03)00346-9}}.

\bibitem{Wulfinghoff2015EfficientApproach}
S.~Wulfinghoff, S.~Reese, {Efficient computational homogenization of simple
  elasto-plastic microstructures using a modified Ritz-Galerkin approach}, in:
  COMPLAS XIII: proceedings of the XIII International Conference on
  Computational Plasticity: fundamentals and applications, CIMNE, 2015, pp.
  956--964.

\bibitem{Kouznetsova2002Multi-scaleScheme}
V.~Kouznetsova, M.~G. Geers, W.~A. Brekelmans, {Multi-scale constitutive
  modelling of heterogeneous materials with a gradient-enhanced computational
  homogenization scheme}, International Journal for Numerical Methods in
  Engineering 54~(8) (2002) 1235--1260.
\newblock \href {http://dx.doi.org/10.1002/nme.541}
  {\path{doi:10.1002/nme.541}}.

\bibitem{Lagzdins1992OrientationalSolids}
A.~Lagzdin{\v{s}}, {Orientational averaging in mechanics of solids}, Vol. 265,
  Longman Scientific and Technical, 1992.

\bibitem{Hill1965ContinuumPolycrystals}
R.~Hill, {Continuum micro-mechanics of elastoplastic polycrystals}, Journal of
  the Mechanics and Physics of Solids 13~(2) (1965) 89--101.
\newblock \href {http://dx.doi.org/10.1016/0022-5096(65)90023-2}
  {\path{doi:10.1016/0022-5096(65)90023-2}}.

\bibitem{Hill1965AMATERIALS}
R.~R. Hill, {A self-consistent mechanics of composite materials}, Mech. Phys.
  Solids 13 (1965) 213--222.

\bibitem{Kroner1961ZurVielkristalls}
E.~Kr{\"{o}}ner, {Zur plastischen verformung des vielkristalls}, Acta
  metallurgica 9~(2) (1961) 155--161.
\newblock \href {http://dx.doi.org/10.1016/0001-6160(61)90060-8}
  {\path{doi:10.1016/0001-6160(61)90060-8}}.

\bibitem{Eshelby1957TheProblems}
J.~D. Eshelby, {The determination of the elastic field of an ellipsoidal
  inclusion, and related problems}, Proc. R. Soc. Lond. A 241~(1226) (1957)
  376--396.
\newblock \href {http://dx.doi.org/10.1098/rspa.1957.0133}
  {\path{doi:10.1098/rspa.1957.0133}}.

\bibitem{PonteCastaneda1991TheComposites}
P.~Ponte~Casta{\~{n}}eda, {The effective mechanical properties of nonlinear
  isotropic composites}, Journal of the Mechanics and Physics of Solids 39~(1)
  (1991) 45--71.
\newblock \href {http://dx.doi.org/10.1016/0022-5096(91)90030-R}
  {\path{doi:10.1016/0022-5096(91)90030-R}}.

\bibitem{PonteCastaneda2002Second-orderTheory}
P.~Ponte~Casta{\~{n}}eda, {Second-order homogenization estimates for nonlinear
  composites incorporating field fluctuations: I - Theory}, Journal of the
  Mechanics and Physics of Solids 50~(4) (2002) 737--757.
\newblock \href {http://dx.doi.org/10.1016/S0022-5096(01)00099-0}
  {\path{doi:10.1016/S0022-5096(01)00099-0}}.

\bibitem{Suquet1993OverallComposites}
P.~M. Suquet, {Overall potentials and extremal surfaces of power law or ideally
  plastic composites}, Journal of the Mechanics and Physics of Solids 41~(6)
  (1993) 981--1002.
\newblock \href {http://dx.doi.org/10.1016/0022-5096(93)90051-G}
  {\path{doi:10.1016/0022-5096(93)90051-G}}.

\bibitem{Bakhvalov1989Homogenisation:Media}
N.~Bakhvalov, G.~Panasenko, {Homogenisation: Averaging Processes in Periodic
  Media: mathematical problems in the mechanics of composite materials},
  Kluwer, Dordrecht, The Netherlands, 1989.
\newblock \href {http://dx.doi.org/10.1007/978-94-009-2247-1}
  {\path{doi:10.1007/978-94-009-2247-1}}.

\bibitem{PentaAnHomogenization}
R.~Penta, A.~Gerisch, {An Introduction to Asymptotic Homogenization}, in:
  A.~Gerisch, R.~Penta, J.~Lang (Eds.), Multiscale Models in Mechano and Tumor
  Biology, Springer, 2018.
\newblock \href {http://dx.doi.org/10.1007/978-3-319-73371-5_1}
  {\path{doi:10.1007/978-3-319-73371-5_1}}.

\bibitem{Bensoussan2011AsymptoticStructures}
A.~Bensoussan, J.-L. Lions, G.~Papanicolaou, {Asymptotic analysis for periodic
  structures}, Vol. 374, American Mathematical Soc., 2011.

\bibitem{Auriault2010HomogenizationMedia}
J.~L. Auriault, C.~Boutin, C.~Geindreau, {Homogenization of Coupled Phenomena
  in Heterogenous Media}, John Wiley {\&} Sons, 2010.
\newblock \href {http://dx.doi.org/10.1002/9780470612033}
  {\path{doi:10.1002/9780470612033}}.

\bibitem{Hori1999OnSolids}
M.~Hori, S.~Nemat-Nasser, {On two micromechanics theories for determining
  micro-macro relations in heterogeneous solids}, Mechanics of Materials 31
  (1999) 667--682.
\newblock \href {http://dx.doi.org/10.1016/S0167-6636(99)00020-4}
  {\path{doi:10.1016/S0167-6636(99)00020-4}}.

\bibitem{Doghri2003HomogenizationAlgorithms}
I.~Doghri, A.~Ouaar, {Homogenization of two-phase elasto-plastic composite
  materials and structures Study of tangent operators, cyclic plasticity and
  numerical algorithms}, International Journal of Solids and Structures 40~(7)
  (2003) 1681--1712.
\newblock \href {http://dx.doi.org/10.1016/S0020-7683(03)00013-1}
  {\path{doi:10.1016/S0020-7683(03)00013-1}}.

\bibitem{Mori1973AverageInclusions}
T.~Mori, K.~Tanaka, {Average stress in matrix and average elastic energy of
  materials with misfitting inclusions}, Acta Metallurgica 21~(5) (1973)
  571--574.
\newblock \href {http://dx.doi.org/10.1016/0001-6160(73)90064-3}
  {\path{doi:10.1016/0001-6160(73)90064-3}}.

\bibitem{Mercier2012ComparisonMaterials}
S.~Mercier, A.~Molinari, S.~Berbenni, M.~Berveiller, {Comparison of different
  homogenization approaches for elastic-viscoplastic materials}, Modelling and
  Simulation in Materials Science and Engineering 20~(2) (2012) 024004.
\newblock \href {http://dx.doi.org/10.1088/0965-0393/20/2/024004}
  {\path{doi:10.1088/0965-0393/20/2/024004}}.

\bibitem{Nemat-Nasser2013Micromechanics:Materials}
S.~Nemat-Nasser, M.~Hori, {Micromechanics: overall properties of heterogeneous
  materials}, Elsevier, 1993.

\bibitem{Nemat-Nasser1999AveragingPlasticity}
S.~Nemat-Nasser, {Averaging theorems in finite deformation plasticity},
  Mechanics of Materials 31~(8) (1999) 493--523.
\newblock \href {http://dx.doi.org/10.1016/S0167-6636(98)00073-8}
  {\path{doi:10.1016/S0167-6636(98)00073-8}}.

\bibitem{Perdahcoglu2011ConstitutiveHomogenization}
E.~S. PerdahcÄ±o{\u{g}}lu, H.~J.~M. Geijselaers, {Constitutive modeling of
  two phase materials using the mean field method for homogenization},
  International Journal of Material Forming 4~(2) (2011) 93--102.
\newblock \href {http://dx.doi.org/10.1007/s12289-010-1007-6}
  {\path{doi:10.1007/s12289-010-1007-6}}.

\bibitem{Kalamkarov2009AsymptoticStructures}
A.~L. Kalamkarov, I.~V. Andrianov, V.~V. Danishevs'kyy, {Asymptotic
  Homogenization of Composite Materials and Structures}, Applied Mechanics
  Reviews 62~(3) (2009) 030802.
\newblock \href {http://dx.doi.org/10.1115/1.3090830}
  {\path{doi:10.1115/1.3090830}}.

\bibitem{Allaire1992HomogenizationConvergence}
G.~Allaire, {Homogenization and two-scale convergence}, SIAM Journal on
  Mathematical Analysis 23~(6) (1992) 1482--1518.
\newblock \href {http://dx.doi.org/10.1137/0523084}
  {\path{doi:10.1137/0523084}}.

\bibitem{Zhikov2000OnApplications}
V.~V. Zhikov, {On an extension of the method of two-scale convergence and its
  applications}, Sbornik: Mathematics 191~(7) (2000) 973.
\newblock \href {http://dx.doi.org/10.1070/SM2000v191n07ABEH000491}
  {\path{doi:10.1070/SM2000v191n07ABEH000491}}.

\bibitem{Fish1997ComputationalPractice}
J.~Fish, K.~Shek, M.~Pandheeradi, M.~S. Shephard, {Computational plasticity for
  composite structures based on mathematical homogenization: Theory and
  practice}, Computer Methods in Applied Mechanics and Engineering 148 (1997)
  53--73.
\newblock \href {http://dx.doi.org/10.1016/S0045-7825(97)00030-3}
  {\path{doi:10.1016/S0045-7825(97)00030-3}}.

\bibitem{Chung2001AsymptoticApplications}
P.~W. Chung, K.~K. Tamma, R.~R. Namburu, {Asymptotic expansion homogenization
  for heterogeneous media: computational issues and applications}, Composites
  Part A: Applied Science and Manufacturing 32~(9) (2001) 1291--1301.
\newblock \href {http://dx.doi.org/10.1016/S1359-835X(01)00100-2}
  {\path{doi:10.1016/S1359-835X(01)00100-2}}.

\bibitem{Pinho-da-Cruz2009AsymptoticModelling}
J.~Pinho-da Cruz, J.~A. Oliveira, F.~Teixeira-Dias, {Asymptotic homogenisation
  in linear elasticity. Part I: Mathematical formulation and finite element
  modelling}, Computational Materials Science 45~(4) (2009) 1073--1080.
\newblock \href {http://dx.doi.org/10.1016/j.commatsci.2009.02.025}
  {\path{doi:10.1016/j.commatsci.2009.02.025}}.

\bibitem{Oliveira2009AsymptoticApplications}
J.~A. Oliveira, J.~Pinho-da Cruz, F.~Teixeira-Dias, {Asymptotic homogenisation
  in linear elasticity. Part II: Finite element procedures and multiscale
  applications}, Computational Materials Science 45~(4) (2009) 1081--1096.
\newblock \href {http://dx.doi.org/10.1016/j.commatsci.2009.01.027}
  {\path{doi:10.1016/j.commatsci.2009.01.027}}.

\bibitem{Jansson1992HomogenizedStructure}
S.~Jansson, {Homogenized nonlinear constitutive properties and local stress
  concentrations for composites with periodic internal structure},
  International Journal of Solids and Structures 29~(17) (1992) 2181--2200.
\newblock \href {http://dx.doi.org/10.1016/0020-7683(92)90065-2}
  {\path{doi:10.1016/0020-7683(92)90065-2}}.

\bibitem{Lopez-Realpozo2008EffectiveConditions}
J.~C. L{\'{o}}pez-Realpozo, R.~Rodr{\'{i}}guez-Ramos, R.~Guinovart-D{\'{i}}az,
  J.~Bravo-Castillero, L.~P. Fern{\'{a}}ndez, F.~J. Sabina, G.~A. Maugin,
  {Effective properties of non-linear elastic laminated composites with perfect
  and imperfect contact conditions}, Mechanics of Advanced Materials and
  Structures 15~(5) (2008) 375--385.
\newblock \href {http://dx.doi.org/10.1080/15376490801977742}
  {\path{doi:10.1080/15376490801977742}}.

\bibitem{Markenscoff2012AsymptoticMicro-cracks}
X.~Markenscoff, C.~Dascalu, {Asymptotic homogenization analysis for damage
  amplification due to singular interaction of micro-cracks}, Journal of the
  Mechanics and Physics of Solids 60~(8) (2012) 1478--1485.
\newblock \href {http://dx.doi.org/10.1016/j.jmps.2012.04.004}
  {\path{doi:10.1016/j.jmps.2012.04.004}}.

\bibitem{Yang2013NonlinearComposites}
Y.~Yang, F.~Y. Ma, C.~H. Lei, Y.~Y. Liu, J.~Y. Li, {Nonlinear asymptotic
  homogenization and the effective behavior of layered thermoelectric
  composites}, Journal of the Mechanics and Physics of Solids 61~(8) (2013)
  1768--1783.
\newblock \href {http://dx.doi.org/10.1016/j.jmps.2013.03.006}
  {\path{doi:10.1016/j.jmps.2013.03.006}}.

\bibitem{Telega2000EffectiveApproximants}
J.~J. Telega, S.~Tokarzewski, A.~Ga{\l}ka, {Effective Conductivity of Nonlinear
  Two-Phase Media: Homogenization and Two-Point Pad{\'{e}} Approximants}, Acta
  Applicandae Mathematicae 61~(1-3) (2000) 295--315.

\bibitem{Devries1989HomogenizationStructures}
F.~Devries, H.~Dumontet, G.~Duvaut, F.~L{\'{e}}n{\'{e}}, {Homogenization and
  damage for composite structures}, International Journal for Numerical Methods
  in Engineering 27~(2) (1989) 285--298.
\newblock \href {http://dx.doi.org/10.1002/nme.1620270206}
  {\path{doi:10.1002/nme.1620270206}}.

\bibitem{Fish1998COMPUTATIONALHOMOGENIZATION}
J.~Fish, Q.~Yu, K.~Shek, {Computational damage mechanics for composite
  materials based on mathematical homogenization}, International Journal for
  Numerical Methods in Engineering 45 (1999) 1657--1679.
\newblock \href
  {http://dx.doi.org/10.1002/(SICI)1097-0207(19990820)45:11<1657::AID-NME648>3.0.CO;2-H}
  {\path{doi:10.1002/(SICI)1097-0207(19990820)45:11<1657::AID-NME648>3.0.CO;2-H}}.

\bibitem{Fish2008MathematicalLoading}
J.~Fish, R.~Fan, {Mathematical homogenization of nonperiodic heterogeneous
  media subjected to large deformation transient loading}, International
  Journal for Numerical Methods in Engineering 76~(7) (2008) 1044--1064.
\newblock \href {http://dx.doi.org/10.1002/nme.2355}
  {\path{doi:10.1002/nme.2355}}.

\bibitem{Fish1995Multi-gridCase}
J.~Fish, V.~Belsky, {Multi-grid method for periodic heterogeneous media Part 2:
  Multiscale modeling and quality control in multidimensional case}, Computer
  Methods in Applied Mechanics and Engineering 126~(1) (1995) 17--38.
\newblock \href {http://dx.doi.org/10.1016/0045-7825(95)00812-F}
  {\path{doi:10.1016/0045-7825(95)00812-F}}.

\bibitem{Ramirez-Torres2018ThreeTissues}
A.~Ram{\'{i}}rez-Torres, R.~Penta, R.~Rodr{\'{i}}guez-Ramos, J.~Merodio, F.~J.
  Sabina, J.~Bravo-Castillero, R.~Guinovart-D{\'{i}}az, L.~Preziosi, A.~Grillo,
  {Three scales asymptotic homogenization and its application to layered
  hierarchical hard tissues}, International Journal of Solids and Structures
  130 (2018) 190--198.
\newblock \href {http://dx.doi.org/10.1016/j.ijsolstr.2017.09.035}
  {\path{doi:10.1016/j.ijsolstr.2017.09.035}}.

\bibitem{Rezakhani2016AsymptoticMaterials}
R.~Rezakhani, G.~Cusatis, {Asymptotic expansion homogenization of discrete
  fine-scale models with rotational degrees of freedom for the simulation of
  quasi-brittle materials}, Journal of the Mechanics and Physics of Solids 88
  (2016) 320--345.
\newblock \href {http://dx.doi.org/10.1016/j.jmps.2016.01.001}
  {\path{doi:10.1016/j.jmps.2016.01.001}}.

\bibitem{Li2017AShale}
W.~Li, R.~Rezakhani, C.~Jin, X.~Zhou, G.~Cusatis, {A multiscale framework for
  the simulation of the anisotropic mechanical behavior of shale},
  International Journal for Numerical and Analytical Methods in Geomechanics
  41~(14) (2017) 1494--1522.
\newblock \href {http://dx.doi.org/10.1002/nag.2684}
  {\path{doi:10.1002/nag.2684}}.

\bibitem{Smyshlyaev2000OnMedia}
V.~P. Smyshlyaev, K.~D. Cherednichenko, {On rigorous derivation of strain
  gradient effects in the overall behaviour of periodic heterogeneous media},
  Journal of the Mechanics and Physics of Solids 48~(6-7) (2000) 1325--1357.
\newblock \href {http://dx.doi.org/10.1016/S0022-5096(99)00090-3}
  {\path{doi:10.1016/S0022-5096(99)00090-3}}.

\bibitem{Fish1995MultigridCase}
J.~Fish, V.~Belsky, {Multigrid method for periodic heterogeneous media Part 1:
  Convergence studies for one-dimensional case}, Computer Methods in Applied
  Mechanics and Engineering 126~(1-2) (1995) 1--16.
\newblock \href {http://dx.doi.org/10.1016/0045-7825(95)00811-E}
  {\path{doi:10.1016/0045-7825(95)00811-E}}.

\bibitem{Triantafyllidis1996TheModels}
N.~Triantafyllidis, S.~Bardenhagen, {The influence of scale size on the
  stability of periodic solids and the role of associated higher order gradient
  continuum models}, Journal of the Mechanics and Physics of Solids 44~(11)
  (1996) 1891--1928.
\newblock \href {http://dx.doi.org/10.1016/0022-5096(96)00047-6}
  {\path{doi:10.1016/0022-5096(96)00047-6}}.

\bibitem{Andrianov2008HigherMaterials}
I.~V. Andrianov, V.~I. Bolshakov, V.~V. Danishevs'kyy, D.~Weichert, {Higher
  order asymptotic homogenization and wave propagation in periodic composite
  materials}, Proceedings of the Royal Society A: Mathematical, Physical and
  Engineering Sciences 464 (2008) 1181--1201.
\newblock \href {http://dx.doi.org/10.1098/rspa.2007.0267}
  {\path{doi:10.1098/rspa.2007.0267}}.

\bibitem{Suquet1987ElementsMechanics}
P.~M. Suquet, {Elements of Homogenization for Inelastic Solid Mechanics}, in:
  E.~Sanchez-Palencia, A.~Zaoui (Eds.), Homogenization techniques for composite
  media, Springer-Verlag, 1987, pp. 193--278.

\bibitem{Pruchnicki1998HomogenizedExpansion}
E.~Pruchnicki, {Homogenized nonlinear constitutive law using fourier series
  expansion}, International Journal of Solids and Structures 35~(16) (1998)
  1895--1913.
\newblock \href {http://dx.doi.org/10.1016/S0020-7683(97)00128-5}
  {\path{doi:10.1016/S0020-7683(97)00128-5}}.

\bibitem{Pruchnicki1994AMaterial}
E.~Pruchnicki, I.~Shahrour, {A macroscopic elastoplastic constitutive law for
  multilayered media: Application to reinforced earth material}, International
  Journal for Numerical and Analytical Methods in Geomechanics 18~(7) (1994)
  507--518.
\newblock \href {http://dx.doi.org/10.1002/nag.1610180705}
  {\path{doi:10.1002/nag.1610180705}}.

\bibitem{Pruchnicki1998HomogenizedMaterial}
E.~Pruchnicki, {Homogenized elastoplastic properties for a partially cohesive
  composite material}, Zeitschrift f{\"{u}}r angewandte Mathematik und Physik
  ZAMP 49~(4) (1998) 568--589.
\newblock \href {http://dx.doi.org/10.1007/s000000050109}
  {\path{doi:10.1007/s000000050109}}.

\bibitem{Ensan2003ASoils}
M.~N. Ensan, I.~Shahrour, {A macroscopic constitutive law for elasto-plastic
  multilayered materials with imperfect interfaces: Application to reinforced
  soils}, Computers and Geotechnics 30 (2003) 339--345.
\newblock \href {http://dx.doi.org/10.1016/S0266-352X(03)00007-7}
  {\path{doi:10.1016/S0266-352X(03)00007-7}}.

\bibitem{Lourenco1996AMaterials}
P.~B. Louren{\c{c}}o, {A matrix formulation for the elastoplastic
  homogenisation of layered materials}, Mechanics of Cohesive--frictional
  Materials: An International Journal on Experiments, Modelling and Computation
  of Materials and Structures 1~(3) (1996) 273--294.
\newblock \href
  {http://dx.doi.org/10.1002/(SICI)1099-1484(199607)1:3<273::AID-CFM14>3.0.CO;2-T}
  {\path{doi:10.1002/(SICI)1099-1484(199607)1:3<273::AID-CFM14>3.0.CO;2-T}}.

\bibitem{Aboudi2003Higher-orderPhases}
J.~Aboudi, M.~J. Pindera, S.~M. Arnold, {Higher-order theory for periodic
  multiphase materials with inelastic phases}, International Journal of
  Plasticity 19~(6) (2003) 805--847.
\newblock \href {http://dx.doi.org/10.1016/S0749-6419(02)00007-4}
  {\path{doi:10.1016/S0749-6419(02)00007-4}}.

\bibitem{Aboudi1982AComposites}
J.~Aboudi, {A continuum theory for fiber-reinforced elastic-viscoplastic
  composites}, International Journal of Engineering Science 20~(5) (1982)
  605--621.
\newblock \href {http://dx.doi.org/10.1016/0020-7225(82)90115-X}
  {\path{doi:10.1016/0020-7225(82)90115-X}}.

\bibitem{Paley1992MicromechanicalModel}
M.~Paley, J.~Aboudi, {Micromechanical analysis of composites by the generalized
  cells model}, Mechanics of materials 14~(2) (1992) 127--139.
\newblock \href {http://dx.doi.org/10.1016/0167-6636(92)90010-B}
  {\path{doi:10.1016/0167-6636(92)90010-B}}.

\bibitem{Covezzi2017HomogenizationTFA}
F.~Covezzi, S.~de~Miranda, S.~Marfia, E.~Sacco, {Homogenization of
  elastic-viscoplastic composites by the Mixed TFA}, Computer Methods in
  Applied Mechanics and Engineering 318 (2017) 701--723.
\newblock \href {http://dx.doi.org/10.1016/j.cma.2017.02.009}
  {\path{doi:10.1016/j.cma.2017.02.009}}.

\bibitem{Fotiu1996OverallComposites}
P.~A. Fotiu, S.~Nemat-Nasser, {Overall properties of elastic-viscoplastic
  periodic composites}, International Journal of Plasticity 12~(2) (1996)
  163--190.
\newblock \href {http://dx.doi.org/10.1016/S0749-6419(96)00002-2}
  {\path{doi:10.1016/S0749-6419(96)00002-2}}.

\bibitem{Walker1994ThermoviscoplasticSubvolumes}
K.~P. Walker, A.~D. Freed, E.~H. Jordan, {Thermoviscoplastic analysis of
  fibrous periodic composites by the use of triangular subvolumes}, Composites
  science and technology 50~(1) (1994) 71--84.
\newblock \href {http://dx.doi.org/10.1016/0266-3538(94)90127-9}
  {\path{doi:10.1016/0266-3538(94)90127-9}}.

\bibitem{Schweizer2011TheHomogenization}
B.~Schweizer, M.~Veneroni, {The needle problem approach to non-periodic
  homogenization}, Networks and Heterogeneous Media 6~(4) (2011) 755--781.
\newblock \href {http://dx.doi.org/10.3934/nhm.2011.6.755}
  {\path{doi:10.3934/nhm.2011.6.755}}.

\bibitem{Heida2016Non-periodicEquations}
M.~Heida, B.~Schweizer, {Non-periodic homogenization of infinitesimal strain
  plasticity equations}, ZAMM Zeitschrift fur Angewandte Mathematik und
  Mechanik 96~(1) (2016) 5--23.
\newblock \href {http://dx.doi.org/10.1002/zamm.201400112}
  {\path{doi:10.1002/zamm.201400112}}.

\bibitem{Visintin2005OnElasto-plasticity}
A.~Visintin, {On homogenization of elasto-plasticity}, in: Journal of Physics.
  Conference Series, Vol.~22, 2005, pp. 222--234.
\newblock \href {http://dx.doi.org/10.1088/1742-6596/22/1/015}
  {\path{doi:10.1088/1742-6596/22/1/015}}.

\bibitem{Schweizer2015HomogenizationMethods}
B.~Schweizer, M.~Veneroni, {Homogenization of plasticity equations with
  two-scale convergence methods}, Applicable Analysis 94~(2) (2015) 375--398.
\newblock \href {http://dx.doi.org/10.1080/00036811.2014.896992}
  {\path{doi:10.1080/00036811.2014.896992}}.

\bibitem{Nesenenko2007HomogenizationViscoplasticity}
S.~Nesenenko, {Homogenization in viscoplasticity}, SIAM Journal on Mathematical
  Analysis 39~(1) (2007) 236--262.
\newblock \href {http://dx.doi.org/10.1137/060655092}
  {\path{doi:10.1137/060655092}}.

\bibitem{Francfort2014OnElasto-plasticity}
G.~Francfort, A.~Giacomini, {On periodic homogenization in perfect
  elasto-plasticity}, Journal of the European Mathematical Society 16~(3)
  (2014) 409--461.
\newblock \href {http://dx.doi.org/10.4171/JEMS/437}
  {\path{doi:10.4171/JEMS/437}}.

\bibitem{Sab1994HomogenizationPlasticity}
K.~Sab, {Homogenization of non-linear random media by a duality method.
  Application to plasticity}, Asymptotic Analysis 9~(4) (1994) 311--336.
\newblock \href {http://dx.doi.org/10.3233/ASY-1994-9402}
  {\path{doi:10.3233/ASY-1994-9402}}.

\bibitem{Ohno2000HomogenizedStructures}
N.~Ohno, X.~Wu, T.~Matsuda, {Homogenized properties of elastic-viscoplastic
  composites with periodic internal structures}, International Journal of
  Mechanical Sciences 42~(8) (2000) 1519--1536.
\newblock \href {http://dx.doi.org/10.1016/S0020-7403(99)00088-0}
  {\path{doi:10.1016/S0020-7403(99)00088-0}}.

\bibitem{Ramirez-Torres2018AnMedia}
A.~Ram{\'{i}}rez-Torres, S.~D. Stefano, A.~Grillo, R.~Rodr{\'{i}}guez-Ramos,
  J.~Merodio, R.~Penta, {An asymptotic homogenization approach to the
  microstructural evolution of heterogeneous media}, nternational Journal of
  Non-Linear Mechanics 106 (2018) 245--257.
\newblock \href {http://dx.doi.org/10.1016/j.ijnonlinmec.2018.06.012}
  {\path{doi:10.1016/j.ijnonlinmec.2018.06.012}}.

\bibitem{Chung2004ALoads}
P.~W. Chung, K.~K. Tamma, R.~R. Namburu, {A computational approach for
  multi-scale analysis of heterogeneous elasto-plastic media subjected to short
  duration loads}, International Journal for Numerical Methods in Engineering
  59~(6) (2004) 825--848.
\newblock \href {http://dx.doi.org/10.1002/nme.880}
  {\path{doi:10.1002/nme.880}}.

\bibitem{Doghri2016FiniteConstituents}
I.~Doghri, M.~I. El~Ghezal, L.~Adam, {Finite strain mean-field homogenization
  of composite materials with hyperelastic-plastic constituents}, International
  Journal of Plasticity 81 (2016) 40--62.
\newblock \href {http://dx.doi.org/10.1016/j.ijplas.2016.01.009}
  {\path{doi:10.1016/j.ijplas.2016.01.009}}.

\bibitem{Danas2008AMedia}
K.~Danas, M.~I. Idiart, P.~Ponte~Casta{\~{n}}eda, {A homogenization-based
  constitutive model for isotropic viscoplastic porous media}, International
  Journal of Solids and Structures 45~(11-12) (2008) 3392--3409.
\newblock \href {http://dx.doi.org/10.1016/j.ijsolstr.2008.02.007}
  {\path{doi:10.1016/j.ijsolstr.2008.02.007}}.

\bibitem{Suquet1997EffectiveComposites}
P.~Suquet, {Effective Properties of Nonlinear Composites}, in: P.~Suquet (Ed.),
  Continuum Micromechanics, Springer Vienna, Vienna, 1997, pp. 197--264.
\newblock \href {http://dx.doi.org/10.1007/978-3-7091-2662-2_4}
  {\path{doi:10.1007/978-3-7091-2662-2_4}}.

\bibitem{Talbot1985VariationalMedia}
D.~R.~S. Talbot, J.~R. Willis, {Variational principles for inhomogeneous
  non-linear media}, IMA Journal of Applied Mathematics 35~(1) (1985) 39--54.
\newblock \href {http://dx.doi.org/10.1093/imamat/35.1.39}
  {\path{doi:10.1093/imamat/35.1.39}}.

\bibitem{PonteCastaneda1996ExactMaterials}
P.~Ponte~Casta{\~{n}}eda, {Exact second-order estimates for the effective
  mechanical properties of nonlinear composite materials}, J. Mech. Phys.
  Solids 44~(6) (1996) 827--862.
\newblock \href {http://dx.doi.org/10.1016/0022-5096(96)00015-4}
  {\path{doi:10.1016/0022-5096(96)00015-4}}.

\bibitem{Agoras2011HomogenizationComposites}
M.~Agoras, P.~Ponte~Casta{\~{n}}eda, {Homogenization estimates for multi-scale
  nonlinear composites}, European Journal of Mechanics, A/Solids 30~(6) (2011)
  828--843.
\newblock \href {http://dx.doi.org/10.1016/j.euromechsol.2011.05.007}
  {\path{doi:10.1016/j.euromechsol.2011.05.007}}.

\bibitem{Chatzigeorgiou2016PeriodicMaterials}
G.~Chatzigeorgiou, N.~Charalambakis, Y.~Chemisky, F.~Meraghni, {Periodic
  homogenization for fully coupled thermomechanical modeling of dissipative
  generalized standard materials}, International Journal of Plasticity 81
  (2016) 18--39.
\newblock \href {http://dx.doi.org/10.1016/j.ijplas.2016.01.013}
  {\path{doi:10.1016/j.ijplas.2016.01.013}}.

\bibitem{Yu2007VariationalMaterials}
W.~Yu, T.~Tang, {Variational asymptotic method for unit cell homogenization of
  periodically heterogeneous materials}, International Journal of Solids and
  Structures 44~(11-12) (2007) 3738--3755.
\newblock \href {http://dx.doi.org/10.1016/j.ijsolstr.2006.10.020}
  {\path{doi:10.1016/j.ijsolstr.2006.10.020}}.

\bibitem{Zhang2015VariationalComposites}
L.~Zhang, W.~Yu, {Variational asymptotic homogenization of elastoplastic
  composites}, Composite Structures 133 (2015) 947--958.
\newblock \href {http://dx.doi.org/10.1016/j.compstruct.2015.07.117}
  {\path{doi:10.1016/j.compstruct.2015.07.117}}.

\bibitem{BouchitteHOMOGENIZATIONDESIGN}
G.~Bouchitte, P.~Suquet, {Homogenization, plasticity and yield design}, in:
  Composite media and homogenization theory, Birkh{\"{a}}user, Boston, 1991,
  pp. 107--133.
\newblock \href {http://dx.doi.org/10.1007/978-1-4684-6787-1_7}
  {\path{doi:10.1007/978-1-4684-6787-1_7}}.

\bibitem{Gluge2016EffectiveMaterials}
R.~Gl{\"{u}}ge, {Effective plastic properties of laminates made of isotropic
  elastic plastic materials}, Composite Structures 149 (2016) 434--443.
\newblock \href {http://dx.doi.org/10.1016/j.compstruct.2016.04.029}
  {\path{doi:10.1016/j.compstruct.2016.04.029}}.

\bibitem{Gluge2017EffectiveMaterials}
R.~Gl{\"{u}}ge, {Effective yield limits of microstructured materials},
  Composite Structures 176 (2017) 496--504.
\newblock \href {http://dx.doi.org/10.1016/j.compstruct.2017.05.051}
  {\path{doi:10.1016/j.compstruct.2017.05.051}}.

\bibitem{Sawicki1981YieldComposites}
A.~Sawicki, {Yield conditions for layered composites}, International Journal of
  Solids and Structures 17~(10) (1981) 969--979.
\newblock \href {http://dx.doi.org/10.1016/0020-7683(81)90035-4}
  {\path{doi:10.1016/0020-7683(81)90035-4}}.

\bibitem{DeBuhan1991AMaterials}
P.~De~Buhan, A.~Taliercio, {A homogenization approach to the yield strength of
  composite materials}, European Journal of Mechanics. A, Solids 10~(2) (1991)
  129--154.

\bibitem{PonteCastaneda1992OnComposites}
P.~Ponte~Casta{\~{n}}eda, G.~deBotton, {On the homogenized yield strength of
  two-phase composites}, Proc. R. Soc. Lond. A 438~(1903) (1992) 419--431.
\newblock \href {http://dx.doi.org/10.1098/rspa.1992.0116}
  {\path{doi:10.1098/rspa.1992.0116}}.

\bibitem{DeBotton1992OnMaterials}
G.~deBotton, P.~Ponte~Casta{\~{n}}eda, {On the ductility of laminated
  materials}, International Journal of Solids and Structures 23~(19) (1992)
  2329--2353.
\newblock \href {http://dx.doi.org/10.1016/0020-7683(92)90219-J}
  {\path{doi:10.1016/0020-7683(92)90219-J}}.

\bibitem{Shen2017ApproximateVoids}
W.~Q. Shen, J.~Zhang, J.~F. Shao, D.~Kondo, {Approximate macroscopic yield
  criteria for Drucker-Prager type solids with spheroidal voids}, International
  Journal of Plasticity 99 (2017) 221--247.
\newblock \href {http://dx.doi.org/10.1016/j.ijplas.2017.09.008}
  {\path{doi:10.1016/j.ijplas.2017.09.008}}.

\bibitem{ElOmri2000Elastic-plasticComposites}
A.~El~Omri, A.~Fennan, F.~Sidoroff, A.~Hihi, {Elastic-plastic homogenization
  for layered composites}, European Journal of Mechanics, A/Solids 19~(4)
  (2000) 585--601.
\newblock \href {http://dx.doi.org/10.1016/S0997-7538(00)00182-0}
  {\path{doi:10.1016/S0997-7538(00)00182-0}}.

\bibitem{He2012HomogenizationResults}
Q.~C. He, Z.~Q. Feng, {Homogenization of layered elastoplastic composites:
  Theoretical results}, International Journal of Non-Linear Mechanics 47~(2)
  (2012) 367--376.
\newblock \href {http://dx.doi.org/10.1016/j.ijnonlinmec.2011.09.018}
  {\path{doi:10.1016/j.ijnonlinmec.2011.09.018}}.

\bibitem{Poulios2018ADeformations}
K.~Poulios, C.~F. Niordson, {A homogenization method for ductile-brittle
  composite laminates at large deformations}, International Journal for
  Numerical Methods in Engineering 113 (2018) 814--833.
\newblock \href {http://dx.doi.org/10.1002/nme.5637}
  {\path{doi:10.1002/nme.5637}}.

\bibitem{borja2001}
R.~I. Borja, R.~A. Regueiro, {Strain localization in frictional materials
  exhibiting displacement jumps}, Computer Methods in Applied Mechanics and
  Engineering 190~(20-21) (2001) 2555--2580.
\newblock \href {http://dx.doi.org/10.1016/S0045-7825(00)00253-X}
  {\path{doi:10.1016/S0045-7825(00)00253-X}}.

\bibitem{simo1993}
J.~C. Simo, J.~Oliver, F.~Armero, {An analysis of strong discontinuities
  induced by strain-softening in rate-independent inelastic solids},
  Computational Mechanics 12~(5) (1993) 277--296.
\newblock \href {http://dx.doi.org/10.1007/BF00372173}
  {\path{doi:10.1007/BF00372173}}.

\bibitem{lene1982}
F.~Lene, D.~Leguillon, {Homogenized constitutive law for a partially cohesive
  composite material}, International Journal of Solids and Structures 18~(5)
  (1982) 443--458.
\newblock \href {http://dx.doi.org/10.1016/0020-7683(82)90082-8}
  {\path{doi:10.1016/0020-7683(82)90082-8}}.

\bibitem{shkoller1994}
S.~Shkoller, A.~Maewal, G.~A. Hegemier, \href{www.jstor.org/stable/43638002}{{A
  dispersive continuum model of jointed media}}, Quarterly of applied
  mathematics 52~(3) (1994) 481--498.
\newline\urlprefix\url{www.jstor.org/stable/43638002}

\bibitem{murakami1989}
H.~Murakami, G.~A. Hegemier, {Development of a nonlinear continuum model for
  wave propagation in joined media: theory for single joint set}, Mechanics of
  Materials 8~(2-3) (1989) 199--218.
\newblock \href {http://dx.doi.org/10.1016/0167-6636(89)90012-4}
  {\path{doi:10.1016/0167-6636(89)90012-4}}.

\bibitem{White2014AnisotropicIntegration}
J.~A. White, {Anisotropic damage of rock joints during cyclic loading:
  Constitutive framework and numerical integration}, International Journal for
  Numerical and Analytical Methods in Geomechanics 38 (2014) 1036--1057.
\newblock \href {http://dx.doi.org/10.1002/nag.2247}
  {\path{doi:10.1002/nag.2247}}.

\bibitem{Borja2013}
R.~I. Borja, {Plasticity}, Springer, 2013.

\bibitem{Tien2006AnRocks}
Y.~M. Tien, M.~C. Kuo, C.~H. Juang, {An experimental investigation of the
  failure mechanism of simulated transversely isotropic rocks}, International
  Journal of Rock Mechanics and Mining Sciences 43~(8) (2006) 1163--1181.
\newblock \href {http://dx.doi.org/10.1016/j.ijrmms.2006.03.011}
  {\path{doi:10.1016/j.ijrmms.2006.03.011}}.

\bibitem{Ambrose2014FailureConditions}
J.~Ambrose, {Failure of Anisotropic Shales under Triaxial Stress Conditions},
  Ph.D. thesis, Imperial College London (2014).

\bibitem{MogiFLOWDruck}
K.~Mogi, {Flow and fracture of rocks under general triaxial compression}, in:
  4th ISRM Congress. International Society for Rock Mechanics and Rock
  Engineering, 1979.

\bibitem{Kwasniewski2007ZACHOWANIEDYLATANCJA}
M.~Kwa{\'{s}}niewski, {Mechanical behaviour of rocks under true triaxial
  compression conditions-Volumetric strain and dilatancy}, Archives of Mining
  Sciences 52~(3) (2007) 409--435.

\bibitem{white2017thermoplasticity}
J.~A. White, A.~K. Burnham, D.~W. Camp, A thermoplasticity model for oil shale,
  Rock Mechanics and Rock Engineering 50~(3) (2017) 677--688.
\newblock \href {http://dx.doi.org/10.1007/s00603-016-0947-7}
  {\path{doi:10.1007/s00603-016-0947-7}}.

\bibitem{Nova1986AnRocks}
R.~Nova, {An extended Cam Clay model for soft anisotropic rocks}, Computers and
  Geotechnics 2~(2) (1986) 69--88.
\newblock \href {http://dx.doi.org/10.1016/0266-352X(86)90005-4}
  {\path{doi:10.1016/0266-352X(86)90005-4}}.

\bibitem{zhao2018}
Y.~Zhao, S.~J. Semnani, Q.~Yin, R.~I. Borja, {On the strength of transversely
  isotropic rocks}, International Journal for Numerical and Analytical Methods
  in Geomechanics 42 (2018) 1917--1934.
\newblock \href {http://dx.doi.org/10.1002/nag.2809}
  {\path{doi:10.1002/nag.2809}}.

\end{thebibliography}

\end{document}